\documentclass[prx,twocolumn,aps,epsf,showpacs,superscriptaddress,longbibliography]{revtex4-1}

\usepackage[dvipsnames]{xcolor}
\usepackage{physics}
\usepackage{amsmath}
\usepackage{amsfonts}
\usepackage{dsfont}
\usepackage[export]{adjustbox}

\usepackage{graphicx}
\begin{document}

\title{Generalized surface multifractality in 2D disordered systems}

\author{Serafim S.~Babkin}
\affiliation{Institute of Science and Technology, Am Campus 1 3400 Klosterneuburg, Austria}

\author{Jonas F.~Karcher}
\affiliation{{Pennsylvania State University, Department of Physics, University Park, Pennsylvania 16802, USA}}
\affiliation{{Institute for Quantum Materials and Technologies, Karlsruhe Institute of Technology, 76021 Karlsruhe, Germany}}
\affiliation{{Institut f\"ur Theorie der Kondensierten Materie, Karlsruhe Institute of Technology, 76128 Karlsruhe, Germany}}

\author{Igor S.~Burmistrov}
\affiliation{{L. D. Landau Institute for Theoretical Physics, acad. Semenova av. 1-a, 142432 Chernogolovka, Russia}}
\author{Alexander D.~Mirlin}
\affiliation{{Institute for Quantum Materials and Technologies, Karlsruhe Institute of Technology, 76021 Karlsruhe, Germany}}
\affiliation{{Institut f\"ur Theorie der Kondensierten Materie, Karlsruhe Institute of Technology, 76128 Karlsruhe, Germany}}

\date{\today}

\begin{abstract}
Recently, a concept of generalized multifractality, which characterizes fluctuations and correlations of critical eigenstates, was introduced and explored for all ten symmetry classes of disordered systems. Here, by using the non-linear sigma-model (NL$\sigma$M) field theory, we extend the theory of generalized multifractality to boundaries of systems at criticality. Our numerical simulations on two-dimensional (2D) systems of symmetry classes A, C, and AII fully confirm the analytical predictions of pure-scaling observables and Weyl symmetry relations between critical exponents of surface generalized multifractality. This demonstrates validity of the NL$\sigma$M for description of 
Anderson-localization critical phenomena not only in the bulk but also on the boundary. The critical exponents strongly violate generalized parabolicity, in analogy with earlier results for the bulk, corroborating the conclusion that the considered Anderson-localization critical points are not described by conformal field theories.
We further derive relations between generalized surface multifractal spectra and linear combinations of Lyapunov exponents of a strip in quasi-one-dimensional geometry, which hold under assumption of invariance with respect to a logarithmic conformal map. Our numerics demonstrate that these relations hold with an excellent accuracy.
Taken together, our results indicate an intriguing situation: the conformal invariance is broken but holds partially at critical points of Anderson localization. 
 
\end{abstract}

\maketitle

\section{Introduction}
\label{sec:intro}

A quantum state in a disordered electronic system can be either localized or delocalized \cite{Anderson58, 50_years_of_localization}. Transitions between these localized and delocalized phases, which can be driven, e.g., by varying the strength of disorder or the energy, are called Anderson transitions \cite{Evers2008}. In a broader sense, Anderson transitions can also occur between two localized phases of different topology. At the critical point of an Anderson transition, wave functions have unusual statistical properties: their moments scale in a non-trivial, power-law way with the size of the system, $L^d\langle\abs{\psi(r)}^{2q}\rangle\propto L^{-\tau_q}$, where $\tau_q$ are independent critical exponents. This property is called multifractality \cite{Evers2008}.


It was discovered \cite{Subramaniam2006, mildenberger2007boundary, subramaniam2008boundary, Evers2008surf,Obuse-Boundary-2008} that the scaling of moments of wave functions at surfaces $S$ at Anderson transitions is characterized by a distinct set of critical exponents $\tau^{(s)}_q$, i.e., $L^{d-1}\langle\abs{\psi(r\in S)}^{2q}\rangle\propto L^{-\tau^{(s)}_q}$, the phenomenon known as surface (or, equivalently, boundary) multifractality. Experimental studies of the Anderson transition usually require transport measurements, which can be performed by attaching leads to the surfaces of the system, making it possible to study directly the surface multifractality. For example, the surface multifractality is expected to be of relevance for experiments on scanning tunneling microscopy of the surface of a magnetic semiconductor Ga$_{1-x}$Mn$_x$As near the three-dimensional metal-insulator transition \cite{richardella2010visualizing}. Furthermore, it was shown that, for a certain range of $q$,  surface effects have a dominant contribution to the multifractality of the entire system (including bulk and surface) \cite{Subramaniam2006}. 

The field-theoretical approach to Anderson localization is based on the nonlinear sigma model (NL$\sigma$M) \cite{Evers2008} formalism. The problem is greatly enriched by the existence of as many as ten symmetry classes of disordered fermionic systems \cite{Altland1997,Zirnbauer1996,Heinzner2005} and by associated topologies. The NL$\sigma$M framework allows one to address various symmetry and topology classes, focussing on key properties of a given universality class. 

The ``conventional" multifractality characterizes the statistics of amplitudes of a single wave function. It was recently recognized that the Anderson-transition criticality implies a much broader pattern of critical correlations that involve several wave functions; the corresponding concept was termed ``generalized multifractality'' \cite{Karcher2021}. Pure-scaling observables of generalized multifractality are characterized by a family of critical exponents $\tau_{\lambda}$, where the multi-index $\lambda=(q_1,q_2,..., q_k)$ labels different observables. 
(Such a construction appeared, although in a restricted sense, already in early works of Wegner in the context of classification of polynomial composite operators in NL$\sigma$M formalism \cite{wegner1979the, wegner1987anomalous1, wegner1987anomalous2}.)
In the last few years, the generalized multifractality was explored analytically for all ten symmetry classes \cite{Gruzberg2013,Karcher2021, Karcher2022, Karcher2022AII, Karcher2023AIII, Karcher2023}.
In particular, pure-scaling observables for all $\lambda$ were constructed by using the NL$\sigma$M approach. 
 These results were verified by numerical simulations at two-dimensional (2D) Anderson transitions of symmetry classes A, C, AIII, AII, D, and DIII
\cite{Karcher2021, Karcher2022, Karcher2022AII, Karcher2023AIII, Karcher2023}. 
Scaling exponents $x_\lambda$ of the field-theory pure-scaling composite operators directly translate to exponents $\tau_\lambda$ of the corresponding eigenfunction observables. (The difference is only in a term representing a linear function of $|\lambda| \equiv q_1 + \ldots + q_k$, with coefficients determined by the spatial dimensionality and the scaling of average density of states.)
Furthermore, the analytical works predicted exact symmetry relations between the exponents $x_\lambda$  of generalized multifractality, which follow from  Weyl symmetries associated with the NL$\sigma$M manifolds. These relations 
(which extend earlier derived relations for the conventional multifractality \cite{Mirlin2006,Gruzberg2011})
have been also confirmed by numerics. In fact, Weyl symmetries---which are exact and highly non-trivial symmetry relations between critical exponents---can serve as a benchmark for numerical simulations.

An important question is whether Anderson-transition critical points are described by conformal field theories (CFT).  It was recently demonstrated \cite{Karcher2021} that, if a 2D Anderson transition is described by a CFT, the critical exponents $x_\lambda$ (and thus also $\tau_{\lambda}$) have a quadratic dependence on the components $q_1,q_2,\ldots$ of multi-index $\lambda$, the property called generalized parabolicity. 
(We refer the reader to Ref.~\cite{Karcher2021} for a detailed discussion of the question, what correlation functions at the Anderson transition could be possibly described by a CFT.)
Remarkably, in the presence of Weyl symmetries, a generalized parabolic spectrum $x_\lambda$ is uniquely fixed (for a given symmetry class), up to a single constant (an overall prefactor). Numerical simulations for 2D Anderson transitions of several symmetry classes showed strong violation of generalized parabolicity  \cite{Karcher2021, Karcher2022,Karcher2022AII, Karcher2023} and thus of conformal invariance. These results were further corroborated by an exact analytical calculation of a subset of generalized-multifractality exponents at spin quantum Hall transition (class C) \cite{Karcher2022}.  Very recently, the results of Ref.~\cite{Karcher2021} were extended to systems of arbitrary spatial dimensionality  \cite{padayasi2023conformal}, with an implication that conformal invariance does not hold also at Anderson transitions in $d >2$ dimensions. 

The goal of this paper is to explore the  generalized multifractality at surfaces of critical systems at Anderson transitions. For this purpose, we extend the 
construction of pure-scaling generalized-multifractality observables to the surface for all ten symmetry classes.
We then perform a numerical study of generalized surface multifractality for 2D models of three symmetry classes: the Ando model \cite{ando1989numerical} in class AII (both metallic phase and metal-insulator transition), the integer quantum Hall (IQH) plateau transition in the U$(1)$ Chalker-Coddington network model in class A and the spin quantum Hall (SQH) transition in the SU$(2)$ version of the network model. 
We analyze the resulting generalized multifractal spectrum, in particular whether it satisfies Weyl symmetries and to which degree it is independent from the bulk spectrum.

The numerical results fully confirm the analytically derived form of pure-scaling observables, thus demonstrating that the NL$\sigma$M theory works also at the boundary of a critical system. As another manifestation of this fact, we find that Weyl symmetry relations hold with a very good accuracy. At the same time, the generalized parabolicity is strongly violated.

We further derive exact relations between surface generalized-multifractality exponents of a 2D system and Lyapunov exponents for a quasi-one-dimensional (quasi-1D) strip, which hold under the assumption that the system is invariant under the exponential map from a strip to a semicircle (generalizing the result obtained in Ref.~\cite{Obuse2010} for conventional multifractality). 
Our numerical results demonstrate that these relations hold with an excellent accuracy, thus providing an indication of the invariance of the critical theory with respect to this specific conformal transformation.

A terminological comment is appropriate at this point. Both terms ``surface multifractality'' and ``boundary multifractality'' were used in previous literature, with a fully identical meaning. In the same way, we use ``surface'' as a term equivalent to ``boundary'' in application to generalized multifractality in the present work. The surface is understood in a generic sense, i.e., as a $(d-1)$-dimensional boundary of a $d$-dimensional system. In particular, for 2D systems the surface is a 1D edge.

The structure of the paper is as follows. 
In Sec.~\ref{sec:analytics} we present the analytical framework of the surface generalized multifractality and derive relations between 2D and quasi-1D systems (valid under the assumption of invariance with respect to the exponential map). Section \ref{sec:numerics} contains results of the numerical analysis for 2D models. Our results are summarized  in Sec.~\ref{sec:summary}, where we also discuss their implications.

\section{Analytical framework}
\label{sec:analytics}

\subsection{Pure-scaling observables and critical exponents}
\label{sec:surfmf}

As discussed in Sec.~\ref{sec:intro}, generalized multifractality is a hallmark of Anderson-transition critical points. Its essence is a power-law scaling of a large family of observables characterizing critical wave functions:
\begin{align}
 & L^{d}\left\langle P_{\lambda}[\psi]\right\rangle \sim L^{-\tau_{\lambda}}.
 \label{eq:P-lambda-scaling}
\end{align}
Here $L$ is the (linear) system size, $P_{\lambda}[\psi]$ is a composite object expressed in terms of wave functions $\psi$ that are close in energy and evaluated at close spatial positions, and $\langle \ldots \rangle$ denotes the disorder averaging. Further, $\lambda$ is a multi-index, $\lambda=\left(q_{1},q_{2},...,q_{k}\right)$, that labels representations of the symmetry group of the NL$\sigma$M. For the case when all $q_i$ are positive integers satisfying $q_1 \ge q_2 \ge \ldots \ge q_k$, the multi-index $\lambda$ corresponds to a conventional Young diagram. The derivation of the pure-scaling eigenfunction observables  $P_{\lambda}[\psi]$ goes \cite{Gruzberg2013,Karcher2021, Karcher2022, Karcher2022AII, Karcher2023} through the NL$\sigma$M pure-scaling composite operators ${\cal P}_\lambda(Q)$. For symmetry reasons, these composite operators are spherical functions on the NL$\sigma$M manifold. Importantly, ${\cal P}_\lambda(Q)$, and thus also $P_{\lambda}[\psi]$,  depend only on symmetry class.
In other words, we know exact expressions for all pure-scaling observables $P_{\lambda}[\psi]$, even for Anderson transitions characterized by strong-coupling fixed points. 
The factor $L^d$ in Eq.~\eqref{eq:P-lambda-scaling} can be equivalently replaced by a summation over sites of the system (in analogy with a conventional definition of the inverse participation ratios). 

We can now generalize these observables defined in the bulk of a $d$-dimensional systems to its $d-1$ dimensional surface. Importantly, the scaling composite operators ${\cal P}_\lambda(Q)$ of the NL$\sigma$M retain their form, which is governed only by the symmetry of the manifold. This applies also to the eigenfunction observables $P_{\lambda}[\psi]$. The only difference is that the coordinates of eigenfunctions entering $P_{\lambda}[\psi]$ should be now taken near the boundary of the system (e.g., within a distance of a few lattice spacings from the boundary in a lattice model). 
We thus have 
\begin{align}
 & L^{d-1}\left\langle P_{\lambda}[\psi]\right\rangle \sim L^{-\tau_{\lambda}^{(s)}},
\end{align}
where the coordinates of the wavefunctions are restricted to a vicinity of the surface. The superscript $s$ indicates that the critical exponents $\tau_{\lambda}^{(s)}$ describe the generalized multifractality at the surface. 

In the field-theoretical language, natural exponents are scaling dimensions of the corresponding NL$\sigma$M  composite  operators at the surface:
\begin{equation}
\langle {\cal P}_\lambda(Q) \rangle \sim 
L^{-x_{\lambda}^{(s)}}\,,
\end{equation}
where $\langle \ldots \rangle$ means the averaging with the sigma-model action.  The relation between the field-theoretical exponents 
$x_{\lambda}^{(s)}$ and the exponents $\tau_{\lambda}^{(s)}$ that can be directly obtained by numerical simulations (as carried out below) reads:
\begin{align}
& x_{\left(q_{1},q_{2},\ldots\right)}^{(s)}=\Delta_{\left(q_{1},q_{2},\ldots\right)}^{(s)}+|\lambda|x_{(1)}^{(s)} \,,
\label{eq:x-lambda-s-relation}
\\
 & \Delta_{\left(q_{1},q_{2},\ldots\right)}^{(s)}=\tau_{\left(q_{1},q_{2},\ldots\right)}^{(s)}-1-d\left(|\lambda|-1\right)-|\lambda|\mu \,.
 \label{eq:Delta-lambda-s-relation}
\end{align}
Here $|\lambda|=q_1+q_2+\ldots + q_k$, $\mu = x_{(1)}^{(s)} - x_{(1)}^{(b)}$, and $x_{(1)}^{(s)}$ and $x_{(1)}^{(b)}$ are exponents that govern the scaling of the local density of states at the surface and in the bulk, respectively. (Whenever appropriate, we will label bulk exponents by a superscript $(b)$, to clearly distinguish them from surface exponents that have the superscript $(s)$.)
Equations \eqref{eq:x-lambda-s-relation} and \eqref{eq:Delta-lambda-s-relation} are generalization of the relations between the conventional surface-multifractality exponents \cite{Subramaniam2006}. 

The explicit form of pure-scaling eigenfunction observables for all symmetry classes is as follows \cite{Karcher2022AII}. The building blocks for the construction are observables with $q_1 = q_2 = \ldots = q_k = 1$ i.e., $\lambda=(1,1, \ldots, 1)$. For classes without a (pseudo-)spin degree of
freedom (classes A, AI, BDI, AIII, and D), they are given by the absolute value squared of a Slater determinant:
\begin{align}
P_{(1,1,\ldots, 1)}[\psi]=\left|\text{det}\left(\begin{array}{ccc}
\psi_{\alpha,r_{1}} & \psi_{\beta,r_{1}} & \ldots\\
\psi_{\alpha,r_{2}} & \psi_{\beta,r_{2}} & \ldots\\
\ldots & \ldots & \ddots
\end{array}\right)\right|^{2}\label{withoutspin}.
\end{align}
Here the indices $\alpha, \beta, \ldots$ label eigenfunctions (all of them close in energy), while $r_i$ are spatial coordinates.
For classes with a (pseudo-)spin degree of freedom (AII, CII, CI, C, DIII) one has instead:
\begin{gather}
P_{(1,1,\ldots)}[\psi] \notag
\\  = \text{det}\left(\begin{array}{ccc|ccc}
\psi_{\alpha\uparrow r_{1}} & \psi_{\beta\uparrow r_{1}} & \ldots & -\psi_{\alpha\downarrow r_{1}}^{*} & -\psi_{\beta\downarrow r_{1}}^{*} & \ldots\\
\psi_{\alpha\uparrow r_{2}} & \psi_{\beta\uparrow r_{2}} & \ldots & -\psi_{\alpha\downarrow r_{2}}^{*} & -\psi_{\beta\downarrow r_{2}}^{*} & \ldots\\
\ldots & \ldots & \ddots & \ldots & \ldots & \ddots\\
\hline \psi_{\alpha\downarrow r_{1}} & \psi_{\beta\downarrow r_{1}} & \ldots & \psi_{\alpha\uparrow r_{1}}^{*} & \psi_{\beta\uparrow r_{1}}^{*} & \ldots\\
\psi_{\alpha\downarrow r_{2}} & \psi_{\beta\downarrow r_{2}} & \ldots & \psi_{\alpha\uparrow r_{2}}^{*} & \psi_{\beta\uparrow r_{2}}^{*} & \ldots\\
\ldots & \ldots & \ddots & \ldots & \ldots & \ddots
\end{array}\right),
\label{withspin}
\end{gather}
where $\uparrow$ and $\downarrow$ refer to the corresponding spin components.

For a generic multi-index $\lambda = \left(q_{1},q_{2},\ldots q_{k}\right)$, the pure-scaling observables  $P_\lambda[\psi]$ is obtained from the above building blocks as follows 
\begin{gather}
 P_{\lambda}[\psi]=\left(P_{\left(1^{1}\right)}[\psi]\right)^{q_{1}-q_{2}}\left(P_{\left(1^{2}\right)}[\psi]\right)^{q_{2}-q_{3}}\notag
 \\
 \times \ldots \times \left(P_{\left(1^{k-1}\right)}[\psi]\right)^{q_{k-1}-q_{k}}\left(P_{\left(1^{k}\right)}[\psi]\right)^{q_{k}},
 \label{eq:abelian} 
\end{gather}
where $(1^m)$ is an abbreviation for $(\underbrace{1, 1, \dots, 1}_{m})$.  It is seen that the observables $P_\lambda[\psi]$ exhibit Abelian fusion; it is inherited from the corresponding property of the appropriately chosen field-theory (NL$\sigma$M) composite operators ${\cal P}_\lambda(Q)$   \cite{Gruzberg2013, bondesan2017gaussian,Karcher2021, Karcher2022, Karcher2022AII, Karcher2023}. Importantly, $q_i$ do not need to be integer or positive here: they can be arbitrary real (or, in fact, even complex) numbers.

All the coordinates $r_i$ are located at the surface, with a distance $ \sim 1$ (ultraviolet scale) between them. One can also extend the definition and consider $P_\lambda[\psi]$ with distances between the points $r_i$ of order $r$, where $1 \ll r \ll L$. This allows one to consider, in addition to scaling with $L$, also a scaling with $r$, see Sec.~\ref{sec:numerics}. 

For three chiral symmetry classes (BDI, AIII, and CII), the above construction holds if one considers observables belonging to a single sublattice. It can also be extended to a broader class of observables, which involve eigenfunctions on both sublattices \cite{Karcher2023}. Such observables (and the associated critical exponents) are labeled by a pair of multi-indices $\lambda,\lambda'$.
The observable $P_{\lambda,\lambda'}[\psi]$ is obtained as a product of $P_{\lambda}[\psi]$ calculated on one sublattice and  $P_{\lambda'}[\psi]$ on the second sublattice.

The Weyl symmetries \cite{Gruzberg2011, Gruzberg2013, Karcher2021, Karcher2022, Karcher2022AII, Karcher2023} are symmetry relations between the exponents of generalized multifractality,
\begin{equation}
x_\lambda = x_{w\lambda}\,, \qquad w \in W \,,
\label{eq:Weyl_sym}
\end{equation}
where $W$ is the Weyl group that acts in the space of multi-indices (weights) $\lambda$. The action of $W$ is generated by two types of transformations: (i) reflections $q_i \to -c_i - q_i$ and (ii) permutations
$q_i \to q_j + (c_j-c_i)/2$ and $q_j \to q_i + (c_i-c_j)/2$.  The Weyl symmetries are exact in symmetry classes A, AI, AII, C, and CI. (Note that all three classes studied numerically in this paper belong to this subset of symmetry classes.) They also hold in symmmetry classes D and DIII if domain walls associated with jumps between two components of the NL$\sigma$M manifold are suppressed. We refer the reader to Ref.~\cite{Karcher2023} for specifics of Weyl symmetries in chiral classes AIII, BDI, and CII.

It was shown in Ref.~\cite{Karcher2021} that, if one {\it assumes} that the critical theory of a 2D Anderson transition is described by a 2D CFT, the exponents $x_{\left(q_{1},q_{2},\ldots, q_k\right)}$ are quadratic functions of the set $\{q_i\}$, the property termed ``generalized parabolicity''. Moreover, in combination with Weyl symmetry, the generalized parabolicity enforces the form
\begin{align}
 & x_{\left(q_{1},q_{2},\ldots, q_k\right)}^{\mathrm{para}}=- b z_\lambda \equiv - b\sum_{i}q_{i}\left(q_{i}+c_{i}\right),
 \label{eq:gener-parab}
\end{align}
with a single parameter $b$ characterizing the whole spectrum of exponents. Here $z_\lambda$ are eigenvalues of the Laplace-Beltrami operator on the NL$\sigma M$ target space (quadratic Casimir invariants). The constants $c_{j}$ (with $j = 1,2,, \ldots$) depend on the symmetry class. They are determined by (the bosonic part of) the half sum of positive roots, $\rho_b = \sum_j c_j e_j$, for the corresponding NL$\sigma M$ (where $e_j$ is the standard basis in the weight space) and are known for all symmetry classes, see, e.g, Ref.~\cite{Gruzberg2013}. In particular, for three symmetry classes that are studied numerically below in this work, one has
\begin{eqnarray}
& c_j = 1 -2j \,, \qquad &\text{class A}, \nonumber \\
& c_j = 3 -4j \,, \qquad &\text{class AII}, \nonumber \\
& c_j = 1 -4j \,, \qquad &\text{class C}.
\label{eq:cj-A-AII-C} 
\end{eqnarray}

\subsection{Connection between 2D generalized-multifractality exponents and quasi-1D Lyapunov exponents}
\label{sec:q1D}

\begin{figure}
    \centering
    \includegraphics[width=.85\linewidth]{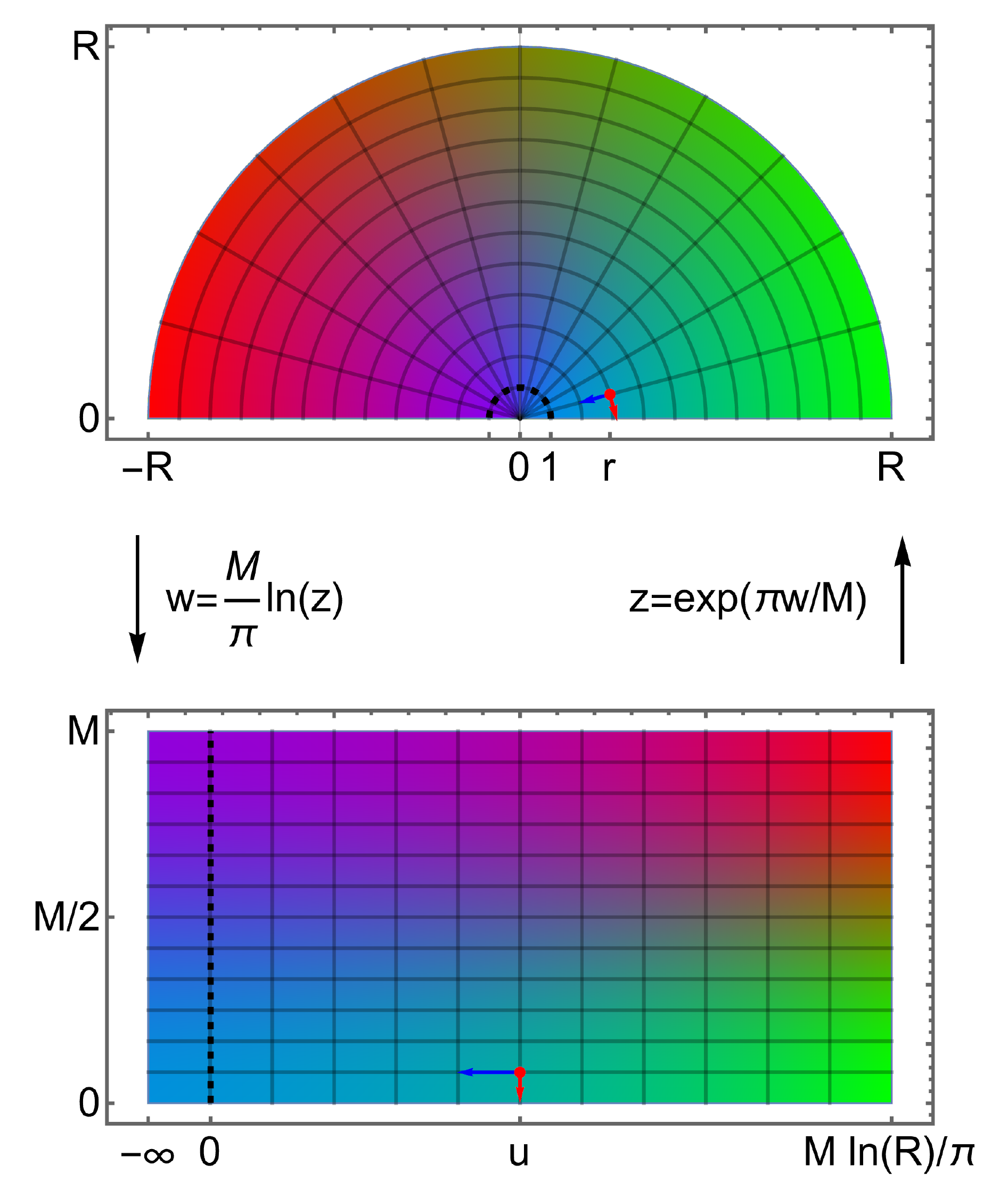}
    \caption{Logarithmic (exponential) conformal mapping \eqref{eq:expm} between a semicircle and an infinite strip with open boundary conditions (on horizontal boundaries). On a lattice there is an ultraviolett cutoff shown as dashed line here at $|z|=1$ or $ u=0 $ respectively. The observation point is located at the real-axis boundary along the real axis of the 2D system; it is shown, as well as its image in the quasi-1D system, by a red dot.}   
    \label{fig:conf}
\end{figure}

We consider the following conformal mapping of a  2D semicircle of radius $R$ to a quasi-1D strip with width $M$ and length $L$:
\begin{align}
 & w=\frac{M}{\pi}\ln z,\qquad z=\exp\left(\frac{\pi}{M}w\right),
 \label{eq:expm}
\end{align}
where $z=x+iy$ is a complex coordinate in the semicircle, with $0 \le |z| \le R$ and $0 \le \arg z \le \pi$, 
and $w=u+iv$ is a complex coordinate in the strip, with
 \begin{align}
 -\infty \le u\le \frac{M}{\pi}\ln R \equiv L \,,\ \ \  0 \le v \le M \,.
 \label{eq:conf2}
\end{align}
This mapping is illustrated in Fig. \ref{fig:conf}.
An ultraviolet cutoff in 2D at $|z|\sim 1$ translates into $u \ge 0$ for the quasi-1D system, yielding a strip of length $L$ with open boundary conditions on horizontal boundaries.   Under the assumption of invariance of the theory with respect to this particular conformal transformation, the NL$\sigma$M for a disordered 2D system defined on a large circle can be mapped on a NL$\sigma$M describing a disordered quasi-1D strip. 

Our analysis generalizes that performed in Ref.~\cite{Obuse2010} for conventional multifractality. We assume that the 2D semicircular system is coupled to a metallic electrode (``absorbing boundary condition") at $|z|=R$, like in Ref.~\cite{Obuse2010}. 
The rest of the boundary (i.e., that on the real axis of $z$) is reflecting.
(Alternatively, the same relation can be derived by placing a small metallic electrode at $R\sim 1$. We briefly comment on this in the end of the derivation.)

We choose an ``observation point'' in the 2D system at some point $r$ on the real-axis boundary, with $1 \ll r \ll R$. The quasi-1D image of this point is at $w = u = (M/\pi) \ln r$, so that $0 < u < L$. The starting point is a relation between averages of operators related  by the conformal transformation formula:
\begin{align}
\langle\mathcal{O}(w)\rangle_{\rm Q1D} &= \left|\dfrac{dz}{dw}\right|^{x_\lambda^{(s)}}\langle\mathcal{O}(z)\rangle_{\rm 2D} 
\label{eq:conf}.
\end{align}
By calculating the scaling of both sides, we will relate the Lyapunov exponents of the quasi-1D system to the 2D generalized-multifractality spectrum. For this purpose, we will consider observables corresponding to $\lambda = (q,q, \ldots, q) \equiv (q^n)$ with $q \to 0$.

In the quasi-1D image, the metallic electrode is attached at $u=L$.
 We need to study how the observable $\langle P_{(q^n)}[\psi] \rangle_{\rm Q1D}$ decays when the coordinate $u$ moves away from $L$. Clearly, in view of the quasi-1D geometry, this decay is exponential as a function of $(L-u)$. We need to find the rate of this exponential decay. 

The wavefunctions $\psi_\alpha$ of the quasi-1D strip are determined by their boundary values $A_\alpha$ at $u =L$ and the transfer matrix $T$ from this boundary to the observation point $u$. We will need $n$ different wave functions that are as close as possible in energy. This amounts to restricting to a fixed energy $E$ for the transfer matrix.
For definiteness, we focus on the spinless situation; generalization to spinful classes is straightforward and leads to the same result. 

All $n$ spatial arguments $w_i$ of wave functions entering $\langle P_{(q^n)}[\psi] \rangle$ are within a distance of order unity from $w=u$. Using the transfer matrix $T_{L-u}$ for a part of the strip of the length $L-u$, we write
\begin{align}
    \psi_{\alpha}(w_{i}) &= B_i T_{L-u} A_\alpha,
\end{align}
where $B_i = P_{v_i} T_{u-u_i}$ selects row $v_i$ of the transfer matrix over $u-u_i$ additional slices and $A_\alpha$ are the initial conditions at $\Re w=L$. Due to the boundary conditions, the columns of $A_\alpha$ correspond to the subspace of wave functions exponentially decaying in the direction of smaller $u$. Without restricting generality, we can choose them to be mutually orthogonal.
We have 
\begin{align}
& P_{(1^n)}[\psi](u)  \nonumber\\
& \simeq \left|\text{det}\left(BP_{n\times n}\right)\text{det}\left(P_{n\times n}T_{L-u}P_{n\times n}\right)\text{det}\left(P_{n\times n}A\right)\right|^2
\label{withoutspinTr},
\end{align}
where the rows of $B$ are given by the $B_i$ and the $A$ is the matrix formed by the $A_\alpha$ as columns. The projection $P_{n\times n}$ restricts to the space of the lowest $n$ Lyapunov exponents. Here  we have neglected exponentially small corrections coming from higher Lyapunov exponents.

Since all $n$ points are distinct by construction, the determinants $\text{det}\left(BP_{n\times n}\right)$ and $\text{det}\left(P_{n\times n}A\right)$ are finite. Importantly, they do not scale with $L-u$ and are therefore of no importance for our purposes. Using Oseledets theorem, we obtain the rate of the exponential decay (in the limit $L-u \to \infty)$
\begin{align}
\langle \ln \text{det}\left(P_{n\times n}T_{L-u}P_{n\times n}\right) \rangle & =  -(\mathcal{L}_1+\ldots+\mathcal{L}_n)(L-u),
\end{align}
where $\mathcal{L}_{1},\mathcal{L}_{2},\ldots,\mathcal{L}_{n}$ are the $n$ smallest Lyapunov
exponents. Note that the Lyapunov exponents describe exponential decay of transfer matrix in a typical realization, which corresponds to averaging the logarithm.
Thus,
\begin{align}
\langle \ln P_{(1^n)}[\psi](u) \rangle_{\rm Q1D} & = -2(\mathcal{L}_1+\ldots+\mathcal{L}_n)(L-u),
\label{ln-P-1n-psi-Q1D}
\end{align}
up to subleading corrections. This argument bears similarity with a discussion of implications of generalized multifractality for transport observables at criticality in Ref.~\cite{Gruzberg2013}.



We turn now to the 2D system, for which we have
\begin{align}
\langle\mathcal{P}_{\lambda}(z)\rangle_{\rm 2D} &
\sim R^{-x_\lambda^{(s)}}\,,
\label{eq:P-lambda-z-2D}
\end{align}
where ${\cal P}_\lambda(z)$ is the field-theory composite operator at the point $z$.
We use the exponential-map relation \eqref{eq:conf}, which involves the factor
\begin{equation}
\left|\dfrac{dz}{dw}\right|^{x_\lambda^{(s)}} =
\left(\dfrac{\pi}{M}\right)^{x_\lambda^{(s)}}|z|^{x_\lambda^{(s)}} = 
\left(\dfrac{\pi}{M}\right)^{x_\lambda^{(s)}}
\exp\left[\dfrac{\pi}{M}x_\lambda^{(s)}u\right].
\label{eq:dzdw}
\end{equation}
Thus, we obtain
\begin{align}
\langle\mathcal{P}_\lambda(w)\rangle_{\rm Q1D}&\sim\left(\dfrac{\pi}{M}\right)^{x_\lambda^{(s)}}\exp\left[-\dfrac{\pi}{M}x_\lambda^{(s)}(L-u)\right].
\label{eq:P-lambda-w-Q1D}
\end{align}
We set now $\lambda = (q^n)$ with small $q$, use 
${\cal P}_{(q^n)} = {[\cal P}_{(1^n)}]^q$, and differentiate both sides of the equation with respect to $q$ at $q=0$. The result reads
\begin{equation}
\langle \ln {\cal P}_{(1^n)} (w)\rangle_{\rm Q1D} = 
- \dfrac{\pi}{M}\dfrac{dx_{(q^n)}^{(s)}}{dq}\Bigg|_{q=0}(L-u).
\label{eq:cmp_2D}
\end{equation}
Equating the exponential decay rate in Eq.~\eqref{eq:cmp_2D} to that in
Eq.~\eqref{ln-P-1n-psi-Q1D} finally yields the sought relation connecting the quasi-1D Lyapunov exponents with 2D generalized-multifractality spectrum:
\begin{align}
 & \pi\frac{dx_{(q^{n})}^{(s)}}{dq}\Bigg|_{q=0}=2M \sum_{i=1}^{n}\mathcal{L}_{i}.
 \label{1D-2D}
\end{align}


As has been pointed out above, an alternative way of deriving this relation is to attach a metallic electrode near the point $z=0$ of the 2D system. We take its radius to be unity. 
In the quasi-1D image, this correspond to attaching a metallic lead at $u=0$.
The observation point $z$ remains unchanged: it is on the real axis of the 2D system, $z=r$, with $1 \ll r \ll R$. Thus, Eq.~\eqref{ln-P-1n-psi-Q1D} becomes
\begin{align}
\langle \ln P_{(1^n)}[\psi] (u) \rangle_{\rm Q1D} & = -2(\mathcal{L}_1+\ldots+\mathcal{L}_n)u,
\label{ln-P-1n-psi-Q1D-alt}
\end{align}
On the 2D side, we have now, instead of 
Eq.~\eqref{eq:P-lambda-z-2D},
\begin{align}
\langle\mathcal{P}_{\lambda}(z)\rangle_{\rm 2D} &
\sim r^{-2x_\lambda^{(s)}}\,.
\end{align}
The factor two in the exponent is because in this case (distance $r$ is much larger than the electrode radius unity) we have effectively a two-point function. (See Ref.~\cite{bondesan2014pure} for a discussion of scaling of bulk correlation functions in the presence of a point-like lead.)
Using the conformal relation \eqref{eq:conf} in combination with Eq.~\eqref{eq:dzdw},
we come to
\begin{align}
\langle\mathcal{P}_\lambda(w)\rangle_{\rm Q1D}&\sim\left(\dfrac{\pi}{M}\right)^{x_\lambda^{(s)}}\exp\left[-\dfrac{\pi}{M}x_\lambda^{(s)}u\right],
\label{eq:P-lambda-w-Q1D-alt}
\end{align}
which is a counterpart of Eq.~\eqref{eq:P-lambda-w-Q1D}. Setting here $\lambda = (q^n)$, taking a derivative with respect to $q$ at $q=0$ and comparing with Eq.~\eqref{ln-P-1n-psi-Q1D-alt}, we come again to the relation  \eqref{1D-2D}.

 The above derivations of the relations \eqref{1D-2D} can be straightforwardly extended to the case of bulk generalized multifractality. In this case, one consider a map of the full circle to a strip with periodic boundary conditions, with $\pi$ replaced by $2\pi$ in Eq.~\eqref{eq:expm}. The result reads
 \begin{align}
 & 2\pi\frac{dx_{(q^{n})}^{(b)}}{dq}\Bigg|_{q=0}=2M \sum_{i=1}^{n}\mathcal{L}_{i}^{(p)} \,,
 \label{1D-2D-bulk}
\end{align}
 where the superscript $(p)$ indicates that the Lyapunov exponents are now calculated for a strip with periodic boundary conditions.

A set of relations \eqref{1D-2D} with $n= 1,2, \ldots$ provides an excellent tool to check whether the critical theory is indeed invariant with respect to the exponential map. Performing such a numerical test for several Anderson-transition critical points is one of the goals of Sec.~\ref{sec:numerics}.

\section{Numerical results}
\label{sec:numerics}

To explore numerically the generalized surface multifractality, we have performed numerical simulations for $L \times L$ critical 2D systems of symmetry class AII (at metal-insulator transition and in the metallic phase, which is ``weakly critical''), class A (IQH transition), and class C (SQH transition). The systems sizes $L$ are in the range $L \in \left[32,...,832\right]$. We perform averaging over $N=10^4$ realizations of disorder as well as over the sample boundary. Critical exponents and their statistical errors are determined by using the procedure described in Ref.~\cite{Karcher2021}.

To test the relations \eqref{1D-2D} (and thus the invariance with respect to the exponential map between quasi-1D and 2D systems), we have also performed the transfer-matrix analysis in quasi-1D geometry. Specifically, we consider very long strips 
(length $L=10^5$)  of width $M$ in the range $\left[32,...,160\right]$. To estimate statistical errors of this approach, an average over 10 disorder configurations is computed. (We recall that Lyapunov exponents are self-averaging in the large-$L$ limit.) For the $n=1$ case of Eq.~\eqref{1D-2D} (corresponding to conventional multifractality), our results below agree with those of Ref.~\cite{Obuse2010}. (Note that our  $2M{\cal L}_1$ corresponds to $2/\Lambda_c$ in notations of Ref.~\cite{Obuse2010}.)  We emphasize, however, that going beyond $n=1$ yields a much more stringent test of the invariance with respect to exponential map. Indeed, as we will see below, the generalized multifractality reveals much stronger violation of generalized parabolicity in comparison with conventional multifractality.

\subsection{Class AII: \ MIT and metallic phase}
\label{sec:aii}

\begin{figure*}[t]
\centering
\includegraphics[width=0.32\textwidth]{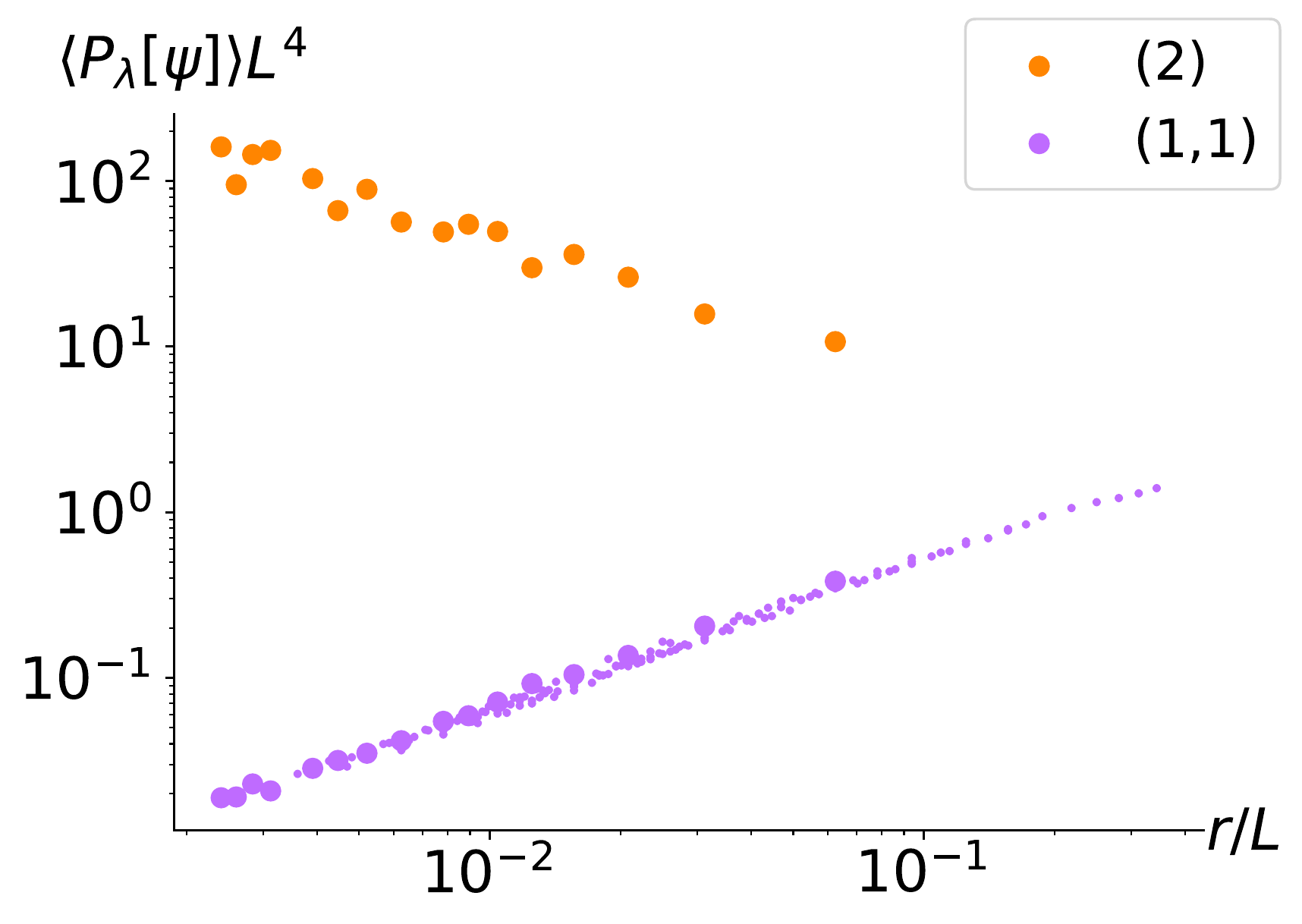}
\includegraphics[width=0.32\textwidth]{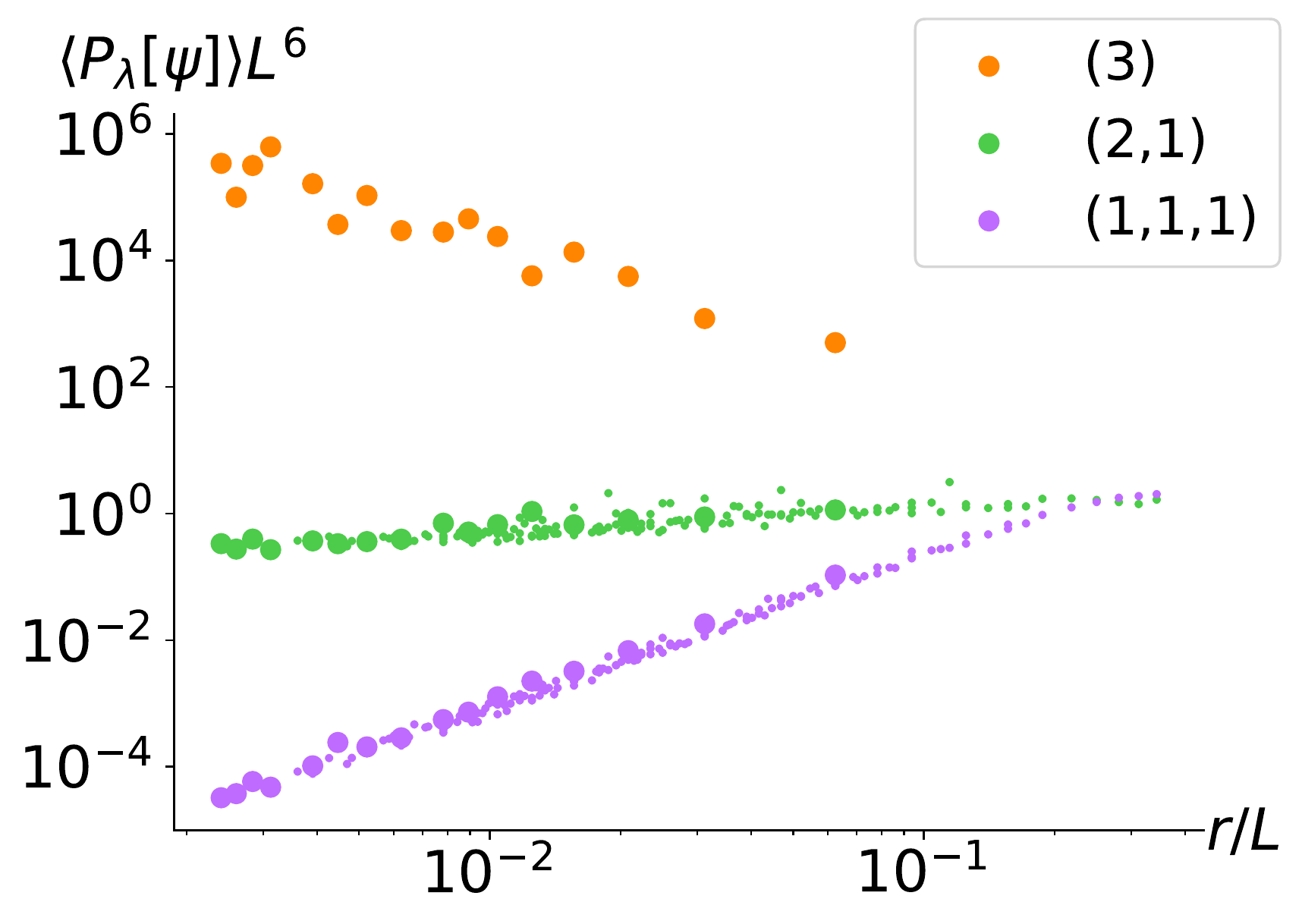}
\includegraphics[width=0.32\textwidth]{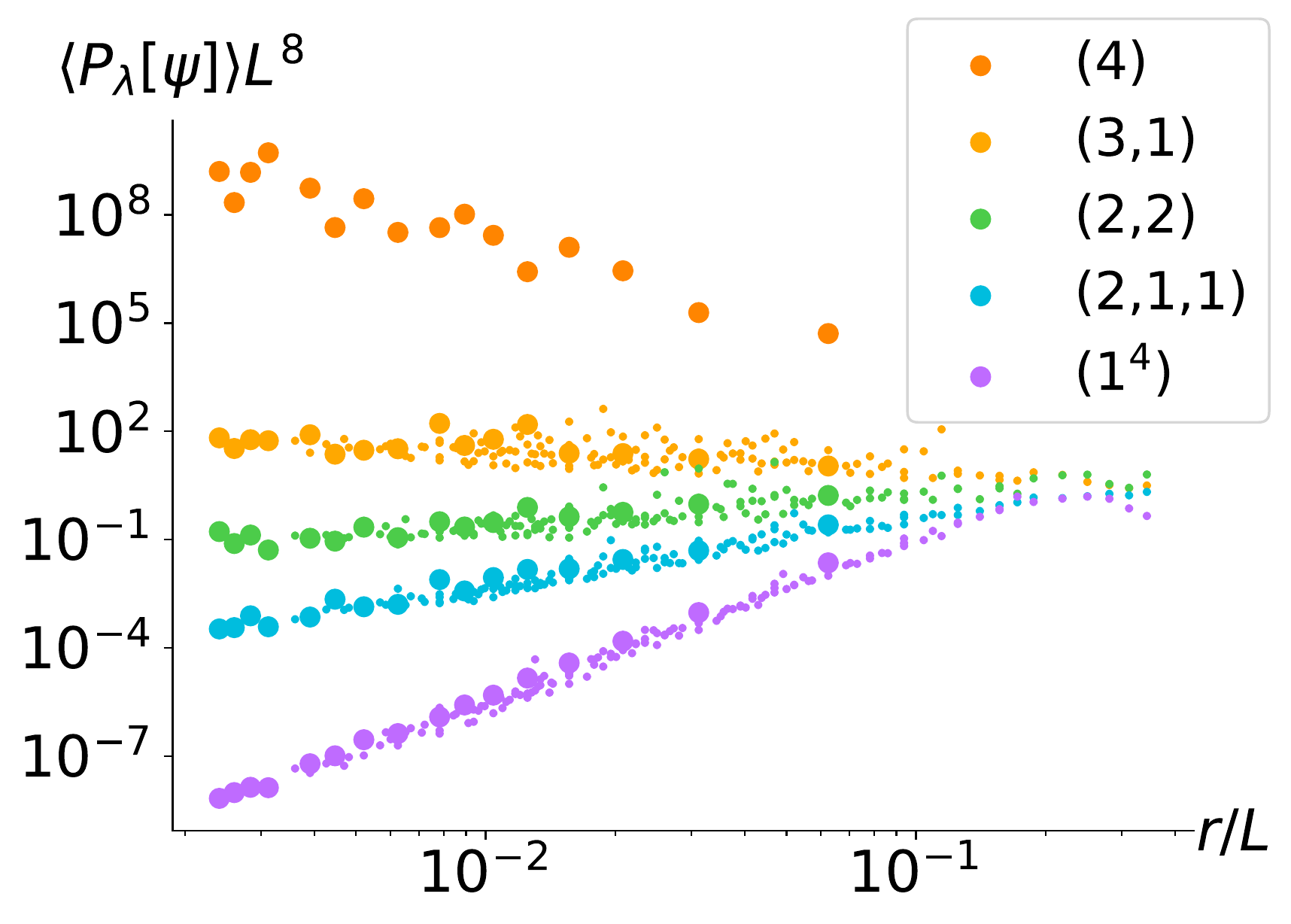}
\caption{Generalized multifractality at MIT in class AII for polynomial observables with $|\lambda|=2$ (left), $3$ (middle), and $4$ (right). The pure-scaling observables $L^{2\left|\lambda\right|} \langle P_{\lambda}[\psi](L,r) \rangle$
are averaged over $N=10^4$ realizations of disorder and over points on the boundary. The data is scaled with  $r^{\Delta_{(q_1)}+...+\Delta_{(q_n)}}$, yielding a collapse as a function of $r/L$ as predicted. Data corresponding to the smallest $r=2$ is highlighted as large dots. (This data was used to extract the exponents as shown in Table~\ref{table:1}.)
}
\label{Fig:AII_MIT_obs}
\end{figure*} 

The symmetry class AII (symplectic Wigner-Dyson class) describes disordered systems with broken spin-rotational symmetry due to a spin-orbit coupling.
This symmetry class exhibits a 2D metal-insulator transition (MIT) \cite{Asada2002,Obuse2007,evers2007wave,Karcher2022AII}. Since this is a spinful symmetry class, the pure-scaling observables have the form \eqref{withspin}. We use the Ando model \cite{ando1989numerical} to study eigenfunction observables numerically. The Hamiltonian is a spinful tight-binding square-lattice model,
\begin{align}
 & H=\sum_{i\sigma}\epsilon_{i}c_{i,\sigma}^{\dagger}c_{i,\sigma}+\sum_{\langle i,j\rangle\sigma\sigma^{\prime}}V_{i,\sigma;j,\sigma^{\prime}}c_{i,\sigma}^{\dagger}c_{j,\sigma^{\prime}},
 \label{Ando_hamiltonian}
\end{align}
where random potentials $\epsilon_{i}$ at each site are drawn uniformly from $\left[-W/2,W/2\right]$, with $i=(x,y)$ labeling 2D lattice sites. The nearest-neighbor hoppings in $x$ and $y$ direction are spin-dependent, $V_{i,\sigma:i+k,\sigma^{\prime}}=V_{0}\exp\left(i\theta_{k}\sigma_{k}\right)$, with $\theta_{x}=\theta_{y}=\pi/6$ and $V_{0}=1$. The defining time-reversal symmetry of this class is realized by $\mathcal{T} = \sigma_2 \mathcal{K}$, where $\mathcal{K}$ stands for complex conjugation. Since we are interested in (generalized) surface multifractality, we introduce a boundary at $y=0$ by setting the hopping $V_{(x,0),\sigma; (x,1),\sigma'}$ to zero for all $x$. 

It is known from previous studies \cite{evers2007wave} that an Anderson MIT takes place in this model at $W_c\approx 5.84$. In addition to studiying the generalized multifractality at this MIT critical point, we also perform numerical simulations for substantially weaker disorder, $W=3$, which puts the system deeply into the metallic phase.

Extracting numerically the generalized-multifractality exponents, we test Weyl-symmetry relations predicted by the NL$\sigma$M. The Weyl symmetry in class AII implies, in particular, the following relations between the exponents corresponding to polynomial observables:
\begin{equation}
    \begin{aligned} & x_{(1)}=0, \quad x_{(1,1)}=x_{(2,2)},\quad x_{(1,1,1)}=x_{(2,2,1)} \,,\\
 & x_{(3,1)}=x_{(2)},\quad x_{(3,2)}=0 \,.
\end{aligned}
\label{Weyl_AII}
\end{equation}
We also study numerically the dependence of critical exponents of the type $x_{(q^n)}$ on $q$. The numerical results allow us to find out whether the generalized-multifractality spectrum satisfies the generalized parabolicity \eqref{eq:gener-parab}. 

It is worth noting that $x_\lambda^{(s)} \equiv \Delta_\lambda^{(s)}$ for class AII, since $x_{(1)}^{(s)}=0$. This holds for all three Wigner-Dyson classes, including class A that studied numerically below in Sec.~\ref{sec:a}.

\begin{table}[h]
\begin{tabular}{|c||c|c|c|}
\hline $\lambda$ & $\tau_{\lambda}^{(s)}$ & $x_{\lambda}^{(s)}$  & $x_{\lambda}^{(s)}/x_{\lambda}^{(b)}$\\
\hline\hline $\left(1\right)$ & 1.005 & $0.005 \pm 0.006$ & - \\
\hline $\left(2\right)$ & 2.18 & -$0.82 \pm 0.06$ & 2.27$\pm$0.17\\
\hline $\left(1,1\right)$ & 3.945 & $0.945 \pm 0.015$ & 1.94$\pm$0.03\\
\hline $\left(3\right)$ & 2.99 & -$2.01 \pm 0.17$ & 1.76$\pm$0.15\\
\hline $\left(2,1\right)$ & 5.46 & $0.46 \pm 0.05$ & 2.04$\pm$0.22\\
\hline $\left(1,1,1\right)$ & 7.48 & $2.48 \pm 0.04$ & 1.86$\pm$ 0.03\\
\hline $\left(4\right)$ & 3.82 & -$3.18 \pm 0.30$ & 1.40$\pm$0.13\\
\hline $\left(3,1\right)$ & 6.63 & -$0.37 \pm 0.18$ & 1.02$\pm$0.50\\
\hline $\left(2,2\right)$ & 7.95 & $0.95 \pm 0.12$ & 1.94$\pm$0.24\\
\hline $\left(2,1,1\right)$ & 9.03 & $2.03 \pm 0.09$ & 1.83$\pm$0.08\\
\hline $\left(1,1,1,1\right)$ & 11.66 & $4.66 \pm 0.05$ & 1.85$\pm$0.02\\
\hline $\left(3,2\right)$ & 9.44 & $0.44 \pm 0.25$ & -\\
\hline $\left(2,2,1\right)$ & 11.39 & $2.39 \pm 0.19$ & 1.80$\pm$0.14\\
\hline \end{tabular}
\caption{Surface generalized-multifractality critical exponents $\tau_{\lambda}^{(s)}$ and $x_\lambda^{(s)}$ at 2D MIT in class AII for all polynomial pure-scaling observables with $\abs{\lambda}\le4$ (and for two observables with $\abs{\lambda}=5$). Statistical error bars (one standard deviation) are shown. The averaging is performed over $2L$ points on the boundary and over $N = 10^4$ realizations
of disorder. The bulk exponents $x^{(b)}_{\lambda}$ are taken from Ref.~\cite{Karcher2022AII}. The column $x_{\lambda}^{(s)}/x_{\lambda}^{(b)}$ demonstrates independence of surface exponents on bulk ones: $x_{\lambda}^{(s)}/x_{\lambda}^{(b)}\ne \text{const}$.
}
\label{table:1}
\end{table}

\subsubsection{Metal-insulator transition}
\label{sec:aii_mit}

\paragraph{Generalized surface multifractality exponents.} To explore the surface generalized multifractality at the MIT, we tune the disorder strength to criticality, $W=5.84\approx W_c$. Numerical results for the dependence of polynomial observables 
$L^{2\left|\lambda\right|}\left\langle P_{\lambda}[\psi]\right\rangle$ with $|\lambda| = 2,3,4$ on $r/L$ are presented in Fig.~\ref{Fig:AII_MIT_obs}. Here $r$ is the point splitting, i.e., the distance between the nearby spatial points entering as arguments of wave functions 
in $P_{\lambda}[\psi]$. The studied values of $r$ are in the range $\{2,3,\ldots,...,11\}$. The data is scaled by the factor $r^{\Delta_{(q_1)}+\ldots+\Delta_{(q_n)}}$, which yields a nice collapse as expected (cf. Ref.~\cite{Karcher2022AII}).  
According to analytical predictions,
\begin{align}
 & L^{2\left|\lambda\right|}\left\langle P_{\lambda}[\psi](L,r)\right\rangle r^{\Delta^{(s)}_{(q_1)}+\ldots+\Delta^{(s)}_{(q_n)}} \sim (L/r)^{-x_{\lambda}^{(s)}}\,,
 \label{power_law_sp}
\end{align}
i.e., when plotted in the log-log representation, the data should represent a fan of straight lines.
This is indeed what is observed in Fig.~\ref{Fig:AII_MIT_obs}. Thus, the prediction of the NL$\sigma$M concerning the form of pure-scaling eigenfunction observables is now verified not only in the bulk but also on the surface. 

 The scaling exponents extracted from the slopes in Fig.~\ref{Fig:AII_MIT_obs} are presented in Table \ref{table:1}. [Also shown are results for two observables corresponding to $|\lambda|=5$, namely, for $\lambda=(3,2)$ and $\lambda=(2,2,1)$.]
More specifically, we use the data for a fixed small $r$ ($r=2$); they are highlighted in Fig.~\ref{Fig:AII_MIT_obs} by larger dots. (An analogous procedure is used for other critical points studied below.) We also present in the table estimated statistical errors (standard deviation). It is seen that these errors increase with the total order $\lambda$ of the operator, which is fully expected. For a given $\lambda$, errors increase with the RG relevance of the operator (quantified by $-x_\lambda^{(s)}$). This is because, with increasing $-x_\lambda^{(s)}$,  the averages are progressively more determined by rare events, which enhances statistical errors. A similar behavior was earlier found for bulk observables. 
 
 We see from the data that all the Weyl symmetry relations \eqref{Weyl_AII} are fulfilled within the statistical uncertainty (for the first three of them, within standard deviation $\sigma$, for the other two within $2\sigma$). The Weyl symmetries between the critical exponents, which were also earlier demonstrated by numerics in the bulk, serve as an excellent confirmation of the validity of the symmetry analysis based on NL$\sigma$M. Thus, we now have this confirmation also for boundary observables. 
 
\begin{figure}[t]
\centerline{\includegraphics[width=0.45\textwidth]{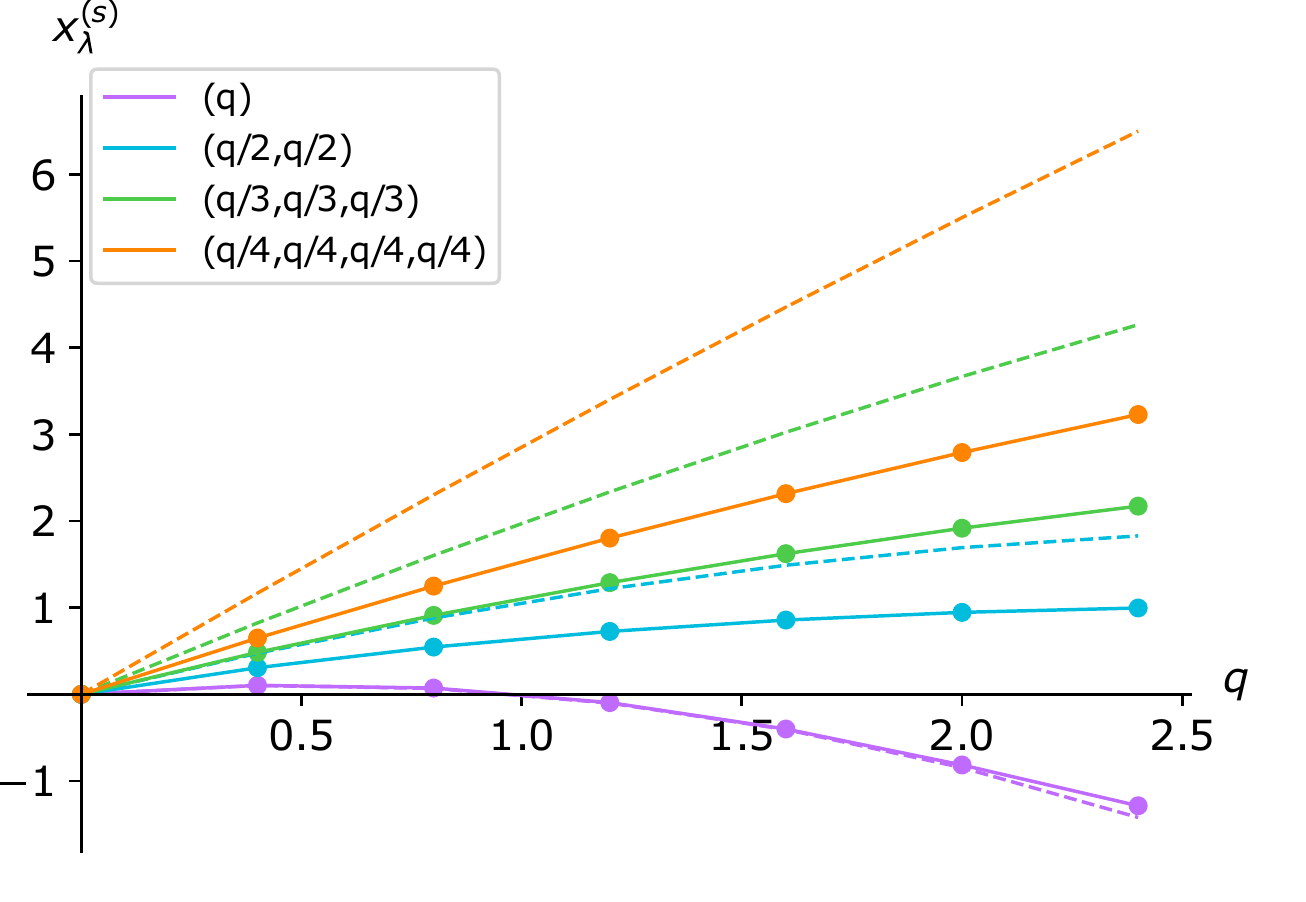}}
\caption{Dependence of critical exponents $x^{(s)}_{\left( (q/n)^n\right)}$ on $q$ for n=1,2,3,4 at MIT in class AII. Solid lines correspond to numerical data, dashed line of the same color -- to generalized parabolicity (\ref{eq:gener-parab}) (with $b=0.423$ chosen in such a way that the parabolic approximation is optimal for $x_{(q)}^{(s)}$ with $0<q<1.6$). It is evident that generalized parabolicity (\ref{eq:gener-parab}) is strongly violated.
} 
\label{Fig:AII_MIT_q_dep}
\end{figure}

 In Fig.~\ref{Fig:AII_MIT_q_dep}, we show the dependence of the critical exponents  $x_{((q/n)^n)}^{(s)}$ for $n\leq 4$ on $q$ and compare them to the prediction from generalized parabolicity (\ref{eq:gener-parab}). (We recall that this is the form of generalized parabolicity for the case when the system obeys Weyl symmetry.)
 The data clearly shows that generalized parabolicity is strongly violated. Indeed, full lines (numerical data) for $n=2$, 3, and 4 differ dramatically from the corresponding generalized-parabolicity curves (shown by dashed lines of the corresponding colors). Crucially, these large deviations exist already for small $q$, which is the region where the accuracy of the numerics is particularly high.

\begin{figure}[t]
\centerline{\includegraphics[width=0.45\textwidth]{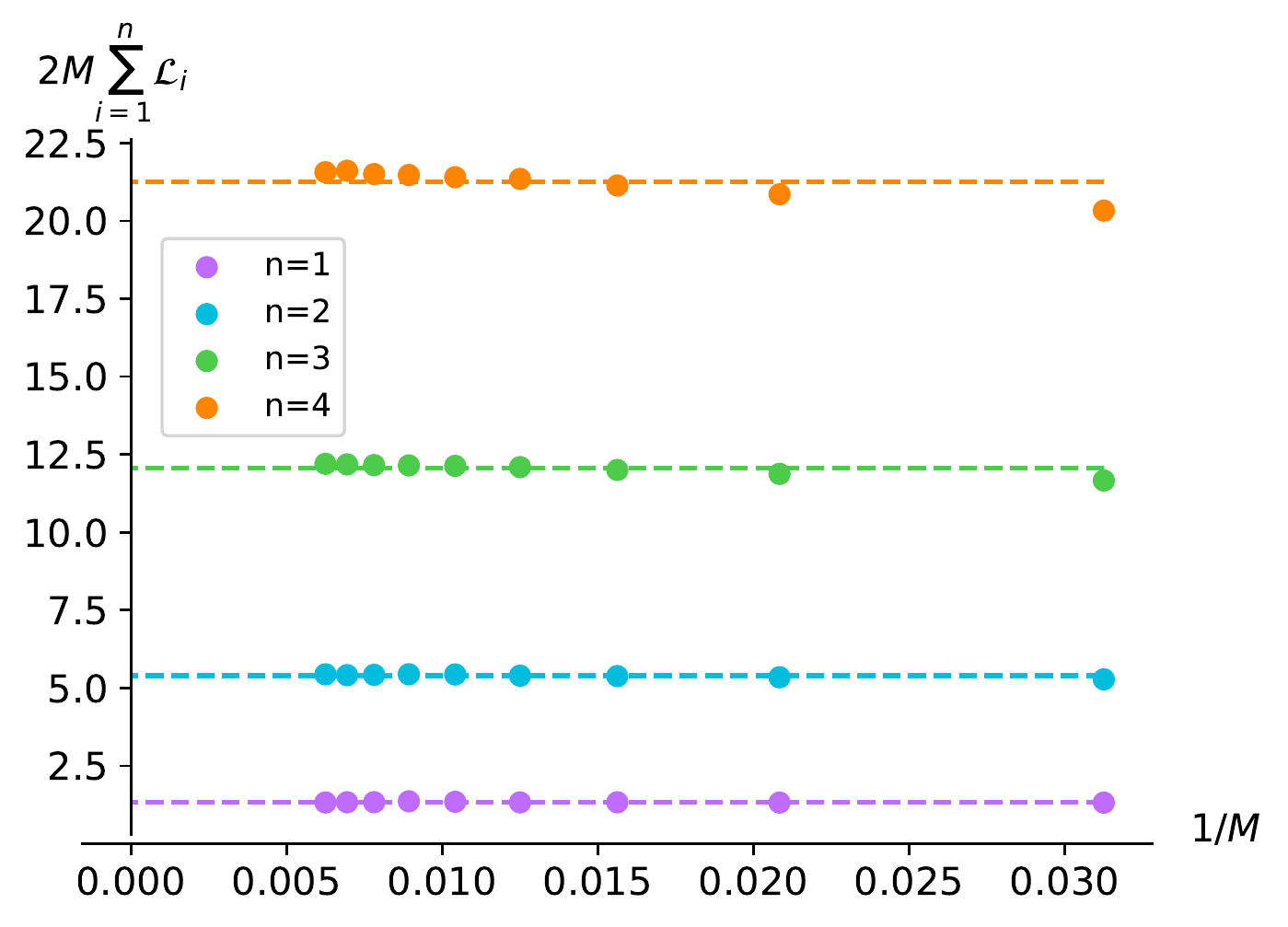}}
\caption{Dependence of Lyapunov exponent sums 2$\sum_{i=1}^{n}\mathcal{L}_i$, with $n=1,2,3,4$,  on the inverse width of the strip $1/M$ for the MIT of class AII. Dashed lines correspond to the values averaged over $M$.  }
\label{Fig:AII_MIT_transfer}
\end{figure} 

\paragraph{Lyapunov exponents.} We turn now to studying numerically whether the equality \eqref{1D-2D} holds. To find the Lyapunov exponents, we implement the transfer matrix approach to the Hamiltonian of the Ando model \eqref{Ando_hamiltonian}. The transfer matrix for the Ando model is derived in the same way as in Ref. \cite{Slevin2014}. Then, we carry out numerical calculations for several values of the strip width $M$. The results for the sums
$2M\sum_{i=1}^{n}\mathcal{L}_{i}$ standing in the right-hand side of Eq.~\eqref{1D-2D}
are shown in Fig.~\ref{Fig:AII_MIT_transfer} for $n =1$, 2, 3, and 4. It is seen that the results are nearly independent on $M$ as they should be for large $M$. Since the $M$-dependence is so weak, we do not perform any extrapolation to $M\to \infty$ but rather average over $M$ (dashed lines in Fig. \ref{Fig:AII_MIT_transfer}).

It is clear from numerical data in Table \ref{table:AII_MIT_comparison} that the equality \eqref{1D-2D} holds with an excellent precision. This demonstrates invariance with respect to the exponential map (at least for the observables whose scaling is described by Eq.~\eqref{1D-2D}).

\begin{table}[h!]
 \begin{tabular}{|c||c|c|}
\hline  & $\pi\frac{dx^{(s)}_{(q^{n})}}{dq}\Bigg|_{q=0}$ & $2M \sum_{i=1}^{n}\mathcal{L}_{i}$\\
\hline\hline $n=1$ & $1.337 \pm 0.020$ & $1.331 \pm 0.005$\\
\hline $n=2$ & $5.42 \pm 0.03$ & $5.39 \pm 0.02$\\
\hline $n=3$ & $12.12 \pm 0.05$ & $12.05 \pm 0.06$\\
\hline $n=4$ & $21.18 \pm 0.07$ & $21.25 \pm 0.14$
\\\hline \end{tabular}
\caption{Numerical results for the expressions on both sides  of Eq.~\eqref{1D-2D} (with $n=1$, 2, 3, and 4) at MIT in class AII: 
the derivative of the exponent $x^{(s)}_{(q^n)}$ and the sum of $n$ lowest Lyapunov exponents $\mathcal{L}_i$.}
\label{table:AII_MIT_comparison}
\end{table}

\begin{table}[h]
\begin{tabular}{|c||c|c|c|c|}
\hline $\lambda$ & $\tau_{\lambda}^{(s)}$ & $x_{\lambda}^{(s)}$ & $2 x_{\lambda}^{(b)}$\\
\hline
\hline $\left(1\right)$ & 0.9988 & -$0.0012 \pm 0.0015$ & - \\
\hline $\left(2\right)$ & 2.890 & -$0.110 \pm 0.004$ & -0.110\\
\hline $\left(1,1\right)$ & 3.206 & $0.206 \pm 0.003$ & 0.219\\
\hline $\left(3\right)$ & 4.667 & -$0.333 \pm 0.010$ & -0.332\\
\hline $\left(2,1\right)$ & 5.095 & $0.095 \pm 0.006$ & 0.109\\
\hline $\left(1,1,1\right)$ & 5.615 & $0.615 \pm 0.007$ & 0.655 \\
\hline $\left(4\right)$ & 6.32 & -$0.68 \pm 0.02$ & -0.67\\
\hline $\left(3,1\right)$ & 6.864 & -$0.136 \pm 0.012$ & -0.111\\
\hline $\left(2,2\right)$ & 7.198 & $0.198 \pm 0.012$ & 0.219\\
\hline $\left(2,1,1\right)$ & 7.503 & $0.503 \pm 0.008$ & 0.5456\\
\hline $\left(1,1,1,1\right)$ & 8.231 & $1.231 \pm 0.009$ & 1.309 \\
\hline $\left(3,2\right)$ & 8.950 & -$0.050 \pm 0.029$ & -0.0012 \\
\hline $\left(2,2,1\right)$ & 9.600 & $0.600 \pm 0.017$ & 0.654 \\
\hline \end{tabular}
\caption{Surface generalized-multifractality critical exponents $\tau_{\lambda}^{(s)}$ and $x_\lambda^{(s)}$ in the metallic phase of Ando model (class AII) for all polynomial pure-scaling observables with $\abs{\lambda}\le4$ (and for two observables with $\abs{\lambda}=5$). Statistical error bars (one standard deviation) are shown. The averaging is performed over $2L$ points on the boundary and over $N = 10^4$ realizations
of disorder. The bulk exponents $x^{(b)}_{\lambda}$ are taken from Ref.~\cite{Karcher2022AII}. 
The analytical (one-loop NL$\sigma$M) prediction $2x^{(b)}_{\lambda} \simeq x^{(s)}_{\lambda}$ holds with a good accuracy.} 
\label{table:AII_metal_obs}
\end{table}

\subsubsection{Metallic phase}
\label{sec:aii_metal}

We consider now the metallic phase in class AII, by choosing the disorder in the Ando model to be $W=3$, i.e., well below $W_c$. This system is not truly at criticality: it flows logarithmically to a ``supermetal'' fixed point with infinite conductivity. However, since this flow is logarithmically slow and, obviously, our numerical study is performed in a limited range of $L$, the system behaves almost as a critical one. 

\paragraph{Generalized surface multifractality.}

The scaling of generalized-multifractality observables
at the surface of a class-AII metallic system is shown in Fig.~\ref{Fig:AII_metal_obs} of Appendix \ref{sec:scal}. The data exhibit fans of almost straight lines, which illustrates a smallnes of logarithmic corrections in this range of $L$. 

The numerically extracted exponents for polynomial observables are collected in Table~\ref{table:AII_metal_obs}. In the limit of large conductance the one-loop NL$\sigma$M calculation predicts that the bulk and surface multifractal exponents are related by a factor of two,
\begin{equation}
x^{(s)}_{\lambda} \simeq 2x^{(b)}_{\lambda},
\label{eq:xs-xb}
\end{equation}
up to parametrically small higher-loop contributions. The exponents presented in the Table~\ref{table:AII_metal_obs} indeed satisfy well this prediction. Further, it is seen that the Weyl symmetry relations \eqref{Weyl_AII} are nicely fulfilled. 

\begin{figure}[t]
\centerline{\includegraphics[width=0.45\textwidth]{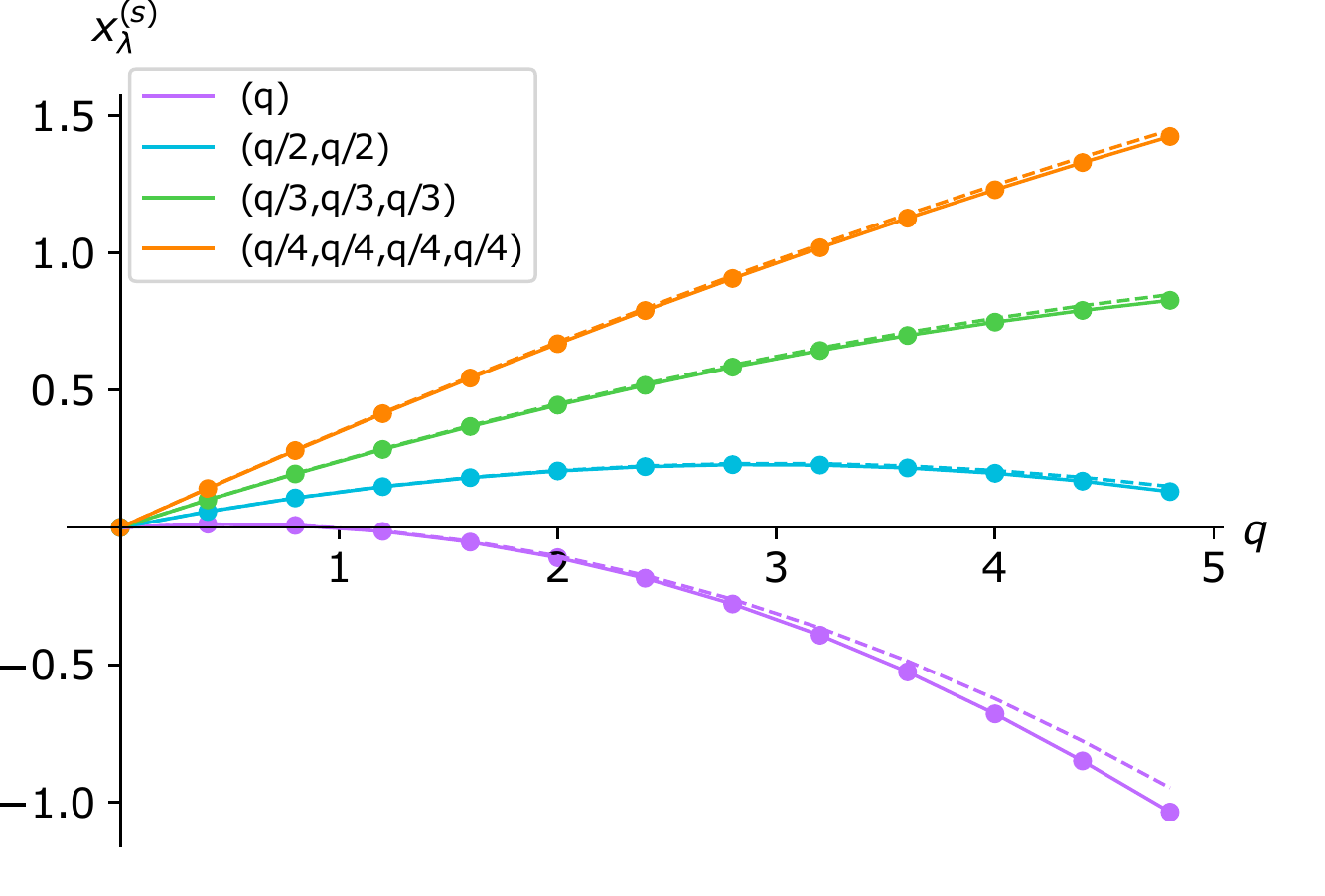}}
\caption{Dependence of critical exponents $x_{\left((q/n)^n\right)}$ on $q$ for n=1,2,3,4 in the metallic phase of Ando model (class AII). Solid lines correspond to numerical data, dashed lines - to generalized parabolicity \eqref{eq:gener-parab} (with $b=0.052$). It is seen that the generalized parabolicity holds with good precision.}
\label{Fig:AII_metal_q_dep}
\end{figure}

The $q$ dependence of critical exponents  $x^{(s)}_{((q/n)^n)}$ for $n\leq 4$ is presented in Fig.~\ref{Fig:AII_metal_q_dep}. In contrast to the MIT critical point (Fig.~\ref{Fig:AII_MIT_q_dep}), generalized parabolicity \eqref{eq:gener-parab} of the spectrum of $x_{\lambda}$ is fulfilled with a good accuracy, in full agrement with analytical predictions based on one-loop NL$\sigma$M analysis.

\begin{figure}[h]
\centerline{\includegraphics[width=0.45\textwidth]{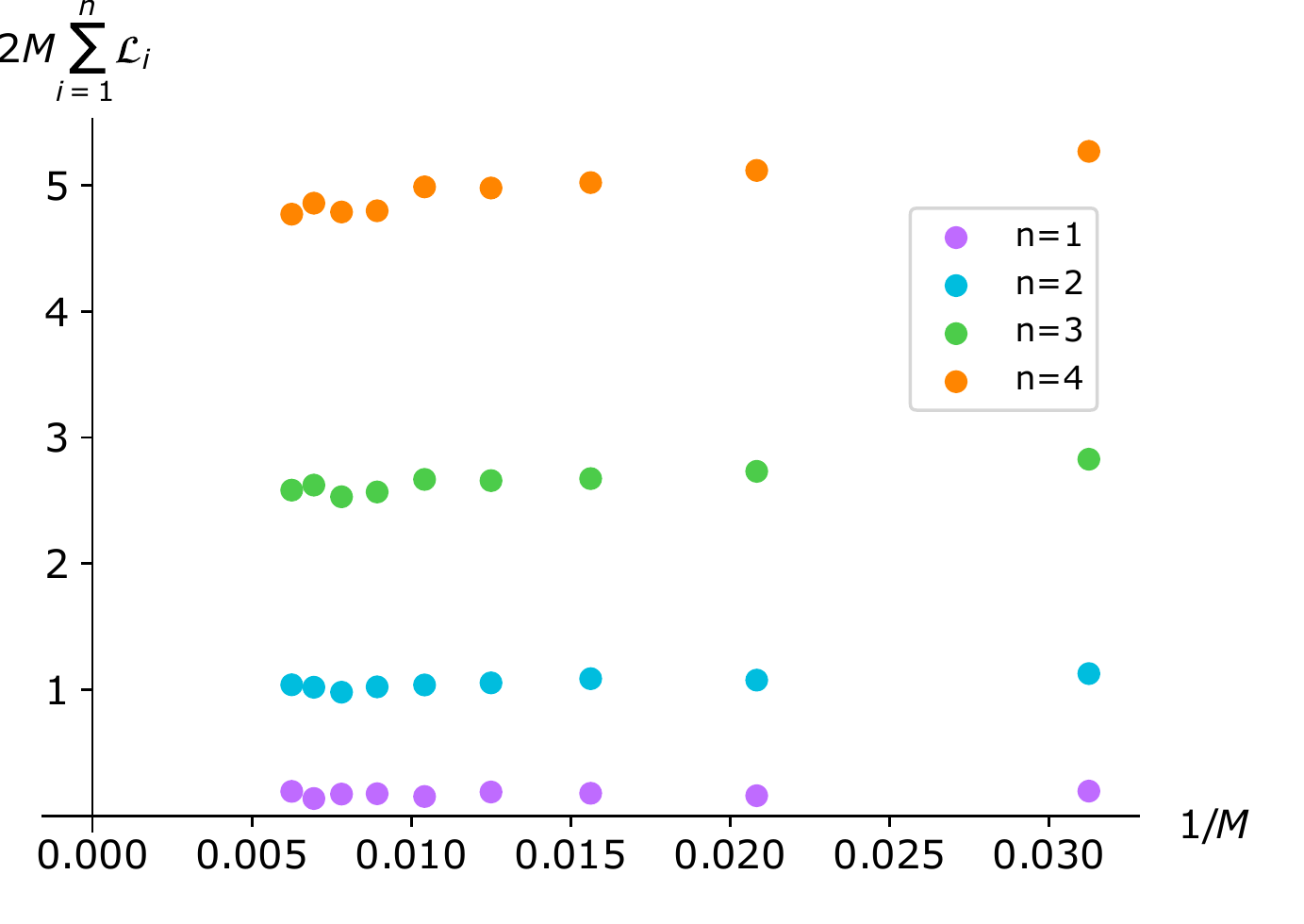}}
\caption{Dependence of Lyapunov exponent sums 2$\sum_{i=1}^{n}\mathcal{L}_i$, with $n=1,2,3,4$,  for  on the inverse width of the strip $1/M$ in the metallic phase of class AII (Ando model with disorder $W=3$). } 
\label{Fig:AII_metal_transfer}
\end{figure} 

\begin{table}[h!]
 \begin{tabular}{|c||c|c|}
\hline  & $\pi\frac{dx_{(q^{n})}^{(s)}}{dq}\Bigg|_{q=0}$ & $2M \sum_{i=1}^{n}\mathcal{L}_{i}$\\
\hline\hline $n=1$ & $0.162 \pm 0.006$ & $0.174 \pm 0.005$\\
\hline $n=2$ & $0.978 \pm 0.007$ & $1.034 \pm 0.007$\\
\hline $n=3$ & $2.434 \pm 0.010$ & $2.565 \pm 0.010$\\
\hline $n=4$ & $4.497 \pm 0.013$ & $4.804 \pm 0.017$
\\\hline \end{tabular}
\caption{Numerical results for the expressions on both sides  of the relation \eqref{1D-2D} (with $n=1$, 2, 3, and 4) in a metallic system of class AII (Ando model with $W=3$):  the derivative of the exponent $x_{(q^n)}^{(s)}$ and the sum of $n$ lowest Lyapunov exponents $\mathcal{L}_i$. The Lyapunov exponents are calculated for $M=160$.}
\label{table:AII_metal_comparison}
\end{table}

\paragraph{Lyapunov exponents.}
In order to find Lyapunov exponents in the AII metal phase and to check numerically the equality \eqref{1D-2D}, we proceed in the same way as for the MIT point in Sec.~\ref{sec:aii_mit}. The results for the sums  2$\sum_{i=1}^{n}\mathcal{L}_i$ of Lyapunov exponents, with $n=1,2,3,4$,  are presented  in Fig.~\ref{Fig:AII_metal_transfer}. A weak logarithmic flow towards smaller values of Lyapunov exponents is visible. Our largest value of the strip width $M=160$, is roughly equal to the geometric mean of the range of $L$ that we use in 2D geometry to extract the critical exponents. We thus take the values of Lyapunov exponents at $M=160$.  In Table \ref{table:AII_metal_comparison} we present numerical values for the expressions on both sides of the relation
\eqref{1D-2D}. One can see that the equality \eqref{1D-2D} holds with good precision as expected. It is worth mentioning that, that, in view of the logarithmic flow discussed above, there is uncertainty in the exact choice of $M$ at which the Lyapunov exponents should be determined. This leads to a relative uncertainty of order $1/\sigma$, where $\sigma$ is 2D conductivity, which is one of possible sources of observed small deviations.

\begin{figure*}[t]
\includegraphics[width=0.45\textwidth]{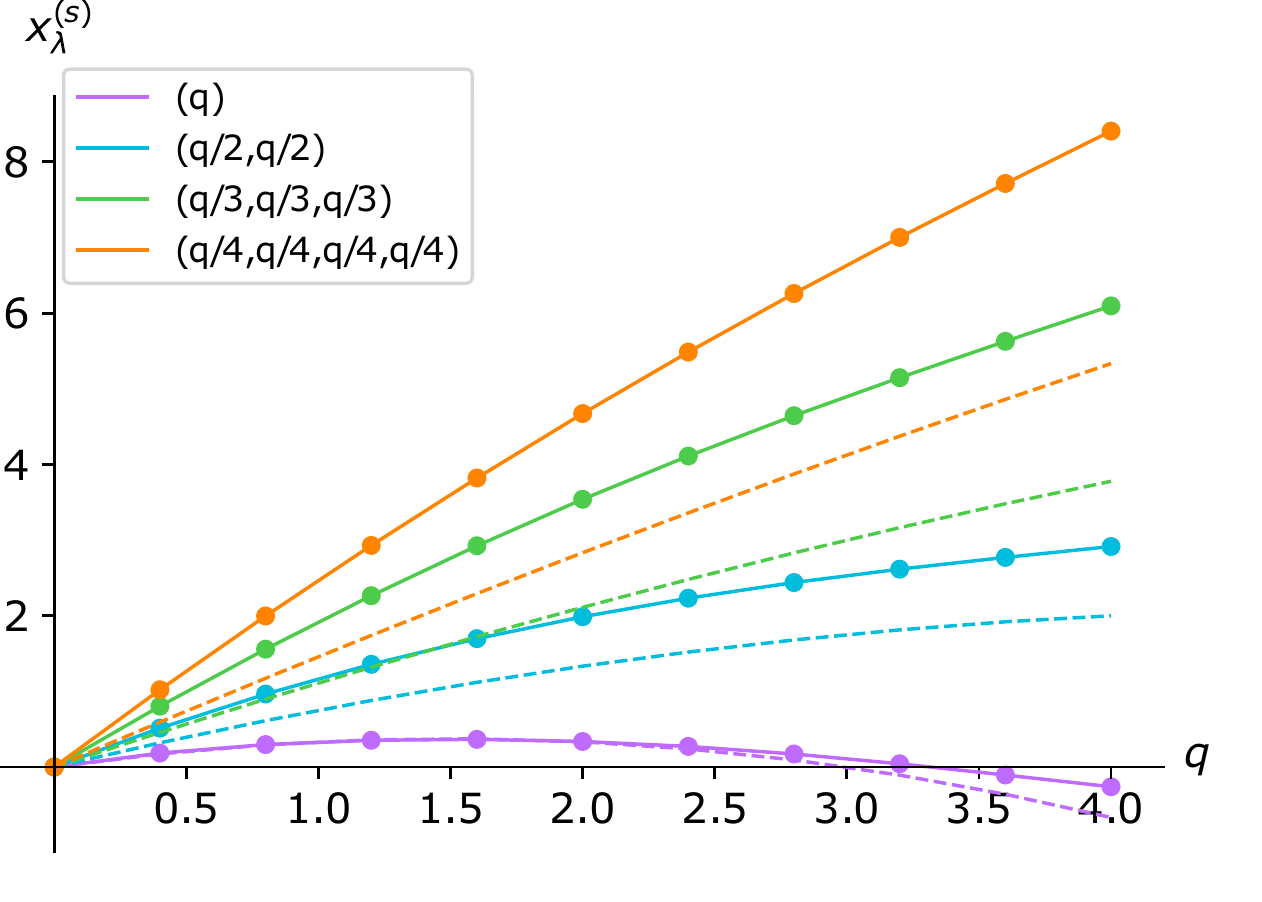}
\includegraphics[width=0.45\textwidth]{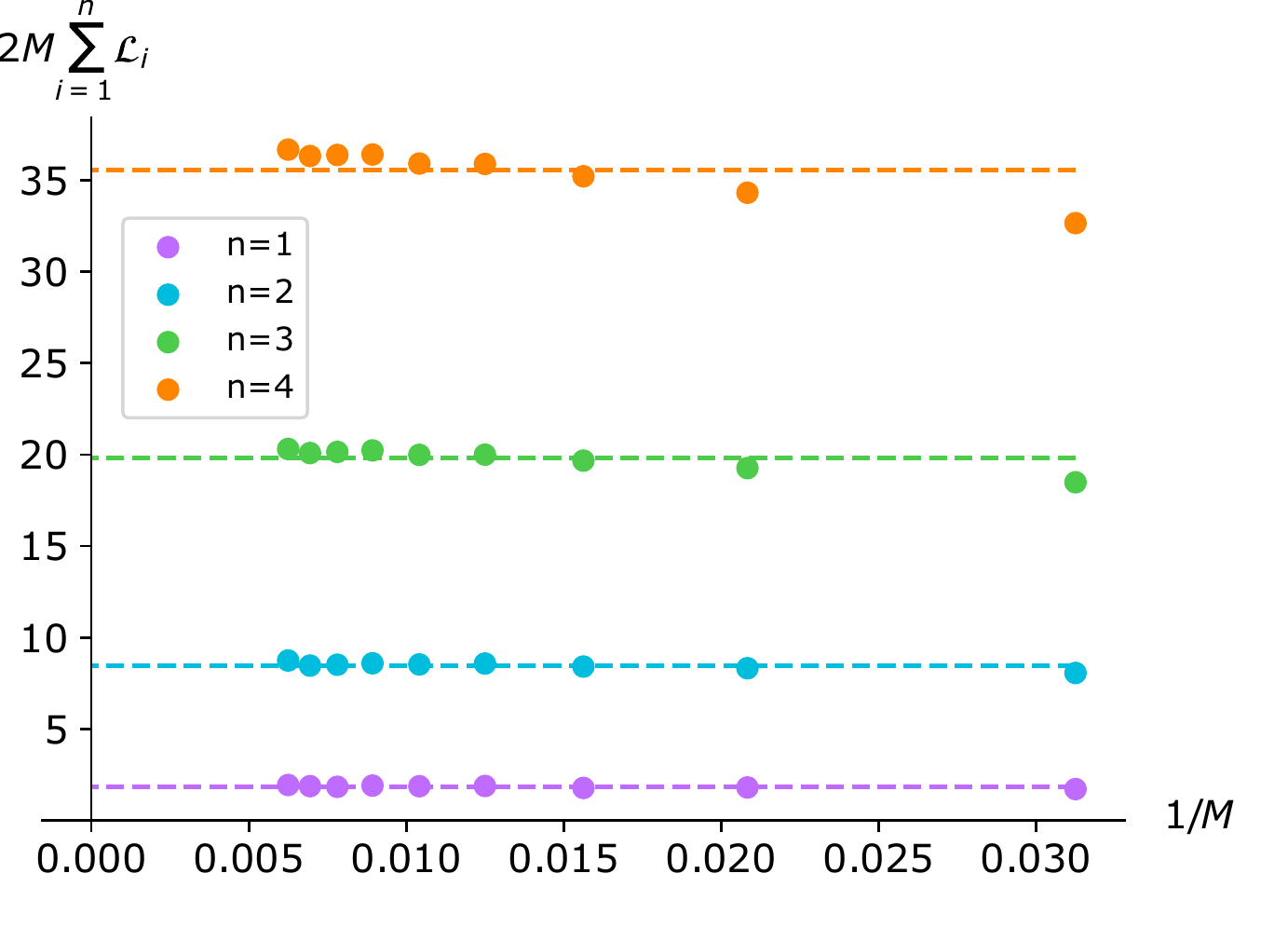}
\includegraphics[width=0.45\textwidth]{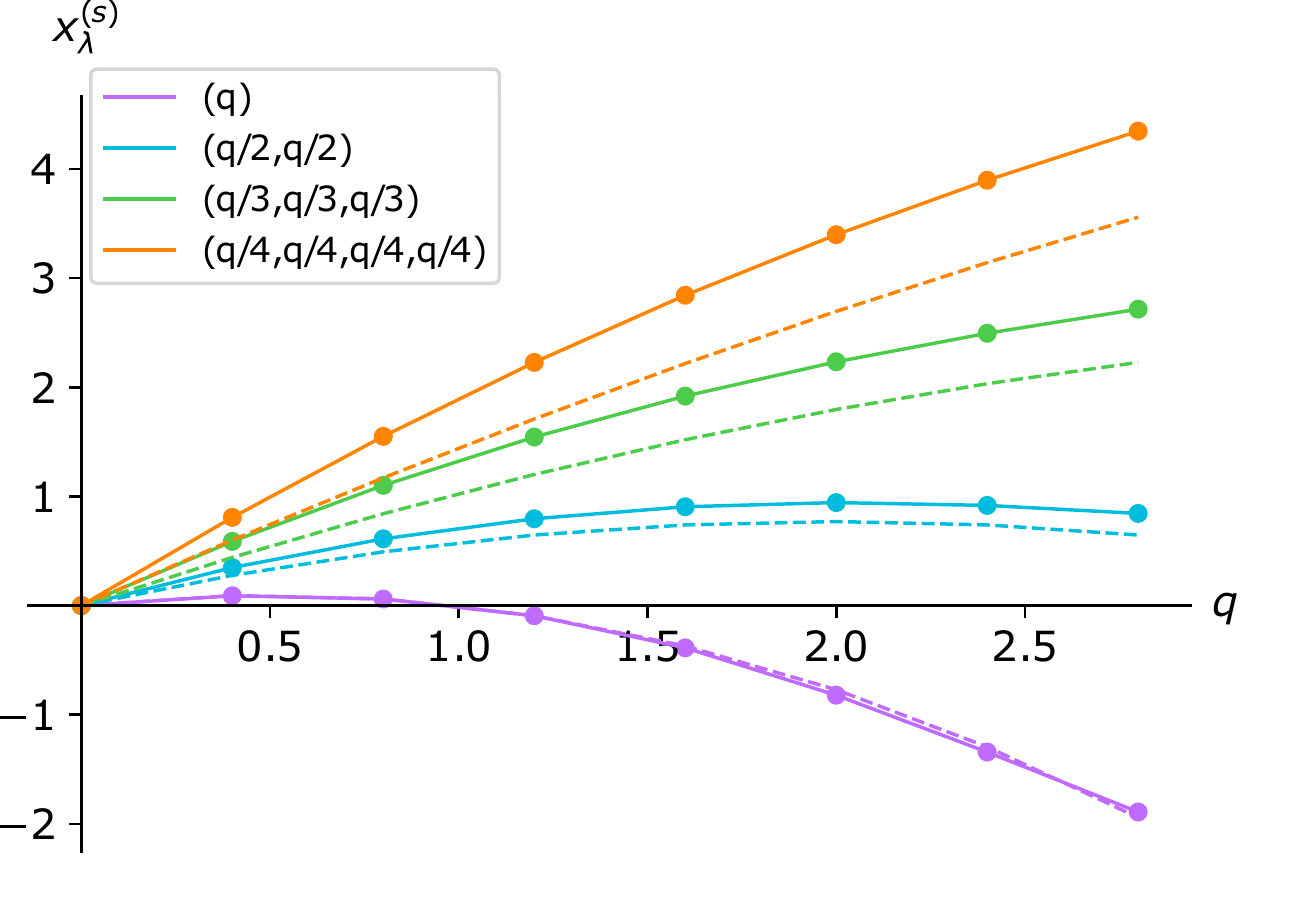}
\includegraphics[width=0.45\textwidth]{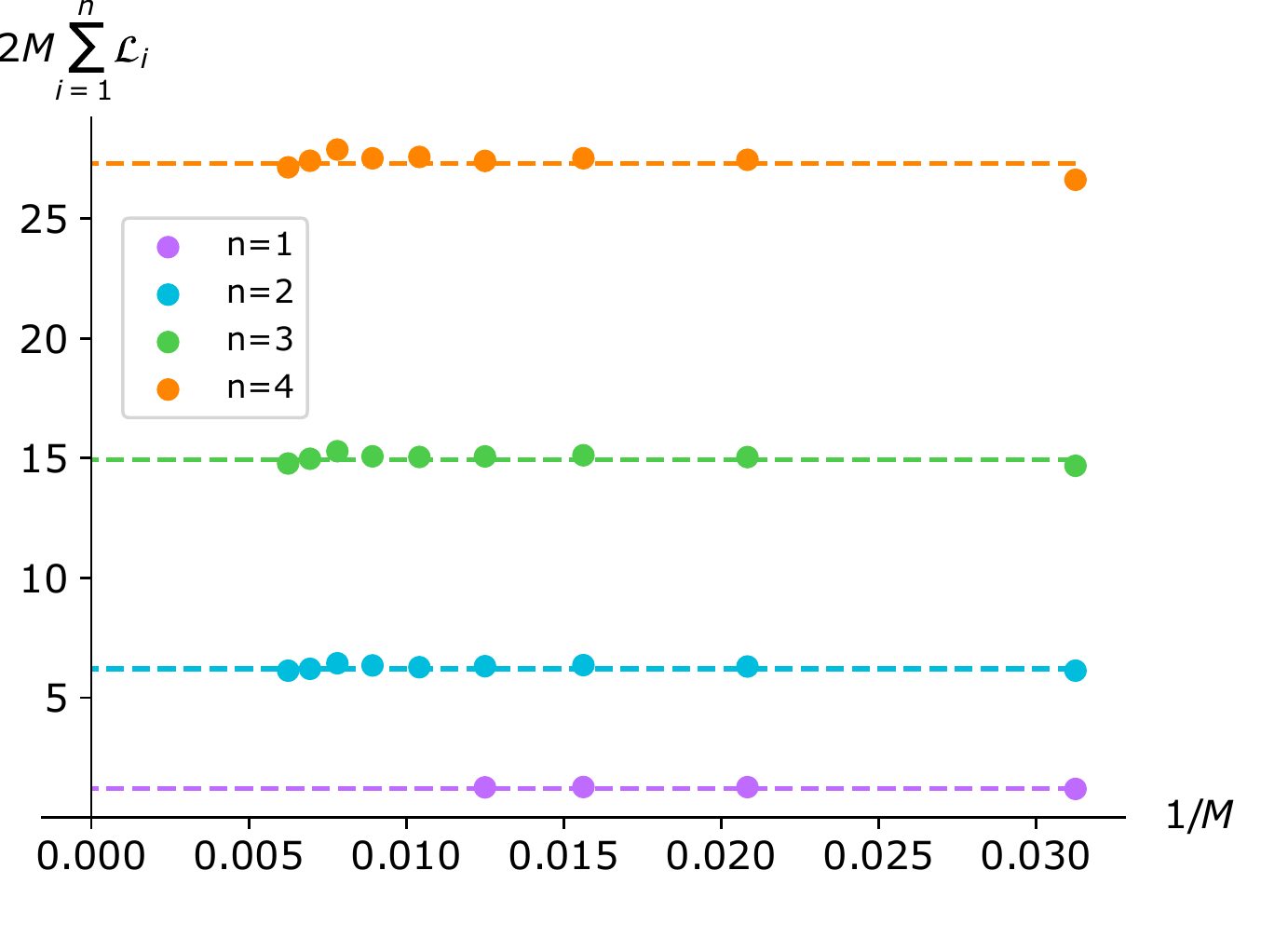}
\caption{Results of network-model simulations for the SQH (top) and IQH (bottom) transitions.
\textit{Upper left panel:} Dependence of critical exponents $x^{(s)}_{\left( (q/n)^n\right)}$ on $q$ for n=1,2,3,4 at SQH transition. Solid lines correspond to numerical data, dashed line of the same color -- to generalized parabolicity (\ref{eq:gener-parab}) with $b=1/6$.
\textit{Upper right panel:} 
Dependence of Lyapunov exponent sums 2$\sum_{i=1}^{n}\mathcal{L}_i$, with $n=1,2,3,4$,   on the inverse width of the strip $1/M$ for the SQH transition. Dashed lines correspond to the values averaged over $M$. 
\textit{Lower panels:} Analogous plots for the IQH transition. The parameter $b$ for the generalized-parabolicity (dashed) lines, Eq.~(\ref{eq:gener-parab}), is $b=0.385$; it is chosen  in such a way that the parabolic approximation is optimal for $x_{(q)}$ with $q<1.6$. 
} 
\label{Fig:C_q_dep}
\end{figure*}

\begin{table}[h!]
 \begin{tabular}{|c||c|c|c|c|c|}
\hline $\lambda$ & $\tau_{\lambda}^{(s)}$ & $\tau_{\lambda,{\rm perc}}^{(s)}$ & $\Delta_{\lambda}^{(s)}$ & $x_{\lambda}^{(s)}$ & $x_{\lambda,{\rm perc}}^{(s)}$\\
\hline\hline $\left(1\right)$ & 1.0815 & 13/12 & -0.0018 & $0.3315 \pm 0.0022$ & 1/3\\
\hline $\left(2\right)$ & 2.838 & 17/6 & -0.329 & $0.338 \pm 0.009$ & 1/3\\
\hline $\left(1,1\right)$ & 4.487 & 4.5 & 1.320 & $1.987 \pm 0.007$ & 2\\
\hline $\left(3\right)$ & 4.36 & 4.25 & -0.89 & $0.11 \pm 0.05$ & 0\\
\hline $\left(2,1\right)$ & 6.27 & 6.25 & 1.02 & $2.02 \pm 0.03$ & 2\\
\hline $\left(1,1,1\right)$ & 9.15 & 9.25 & 3.90 & $4.90 \pm 0.04$ & 5\\
\hline $\left(4\right)$ & 5.74 & - & -1.59 & -$0.26 \pm 0.18$ & -\\
\hline $\left(3,1\right)$ & 7.86 & - & 0.52 & $1.86 \pm 0.09$ & -\\
\hline $\left(2,2\right)$ & 8.92 & - & 1.58 & $2.92 \pm 0.05$ & -\\
\hline $\left(2,1,1\right)$ & 10.83 & 11 & 3.49 & $4.83 \pm 0.12$ & 5\\
\hline $\left(1,1,1,1\right)$ & 14.41 & 46/3 & 7.07 & $8.41 \pm 0.06$ & 28/3
\\\hline \end{tabular}
\caption
{Surface generalized-multifractality critical exponents $\tau_{\lambda}^{(s)}$,  $\Delta_{\lambda}^{(s)}$, and $x_\lambda^{(s)}$ at the SQH transition (class C) for all polynomial pure-scaling observables with $\abs{\lambda}\le4$ . Statistical error bars (one standard deviation) are shown for $x_\lambda^{(s)}$. The averaging is performed over $2L$ points on the boundary and over $N = 10^4$ realizations
of disorder. Also included are exact analytical results 
$x^{(s)}_{\lambda,{\rm perc}}$ and $\tau^{(s)}_{\lambda,{\rm perc}}$ obtained by percolation mapping (for those $\lambda$, for which the mapping is available). 
}
\label{table:C_obs}  
\end{table}

\subsection{Class C: \ SQH transition}
\label{sec:c}

We turn now to the SQH transition, which is a transition between two topologically distinct phases of a 2D system belonging to class C. It can be viewed as a ``superconducting cousin'' of the IQH transition. 
A remarkable property of the SQH transition is that a certain subset of critical exponents can be calculated exactly by a mapping to a classical percolation. This yields a possibility to benchmark numerics and provides a stringent test to any candidate theory of the SQH critical point. 

\subsubsection{Surface generalized multifractality exponents: Analytical results}

The mapping to percolation was first constructed for the average local density of states (LDOS) and average two-point conductance
\cite{Gruzberg1999,Beamond2002}; this yielded, in particular, exact values of the localization length exponent $\nu=4/3$ and the exponent $x^{(b)}_{(1)} = 1/4$ controlling the scaling of (bulk) average LDOS.
It was subsequently demonstrated in Refs.~\cite{evers2003multifractality,mirlin2003wavefunction} that the mapping to percolation can be also developed for the multifractality observables $P_{(q)}$ with $q=2$ and $q=3$, yielding exact values $x_{(2)}^{(b)} = 1/4$ and $x_{(3)}^{(b)}=0$. In Ref.~\cite{subramaniam2008boundary}, this mapping was further extended to the corresponding surface observables, with the results $x^{(s)}_{(1)}= x^{(s)}_{(2)}= 1/3$ and $x^{(s)}_{(3)}= 0$.
More recently, it was discovered \cite{Karcher2022} that the mapping can be constructed to an infinite series of exponents of generalized multifractality.
Specifically, it was shown in Ref.~\cite{Karcher2022} that the mapping to percolation can be constructed for the observables $P_{(1^q)}[\psi]$ with any positive integer $q$. As a result, the  corresponding exponents $x_{(1^q)}$ can be expressed in terms of scaling dimensions  $ x_q^{\rm h} $ of $q$-hull operators of the percolation theory:
\begin{equation}
x_{(1^q)} = x_q^{\rm h} \,.
\end{equation}
The (bulk) $q$-hull exponents of 2D percolation have been found analytically in Ref.~\cite{saleur1987exact},
\begin{equation}
x_q^{{\rm h}(b)} = \frac{4q^2-1}{12} \,, \qquad q= 1, 2, \ldots.
\end{equation}
Thus, Ref.~\cite{Karcher2022} obtained exact analytical predictions for the (bulk) generalized-multifractality exponents $x^{(b)}_{(1^q)}$, which were also confirmed by numerical simulations for $q \le 5$. 

We extend now the mapping of Ref.~\cite{Karcher2022} to the observables $P_{(1^q)}[\psi]$ at the boundary. This yields for the corresponding exponents:
\begin{equation}
x_{(1^q)}^{(s)} = x_q^{{\rm h}(s)} \,,
\end{equation}
where $x_q^{{\rm h}(s)}$ are surface counterparts of $q$-hull percolation exponents. 
The exponents $x_q^{{\rm h}(s)}$ are known
\cite{saleur1989on} ( see Eq.~(3.12) there with $\mu=2$ for the percolation problem and $S \mapsto q$; see also Ref.~\cite{Gruzberg1999}):
\begin{equation}
x_q^{{\rm h}(s)} = \frac{q(2q-1)}{3} \,, \qquad q= 1, 2, \ldots.
\end{equation}
Thus, we obtain exact analytical results for the surface generalized-multifractality exponents $x_{(1^q)}^{(s)}$ at the SQH transition.  Substituting them in Eqs.~\eqref{eq:x-lambda-s-relation}, \eqref{eq:Delta-lambda-s-relation} and using
\begin{equation}
\mu = x_{(1)}^{(s)} - x_{(1)}^{(b)} = \frac13 - \frac14 = \frac{1}{12} \,,
\end{equation}
we also find exact values of the exponents  $\tau_{(1^q)}^{(s)}$,
\begin{equation}
\tau_{(1^q)}^{(s)} = \frac{8q^2 + 17 q - 12}{12} \,, \qquad q= 1, 2, \ldots.
\end{equation}

The Weyl symmetry relations \eqref{eq:Weyl_sym} in class C imply, in particular,
\begin{align}
x_{(3)} = 0 \,, \qquad  x_{(2,1^m)} = x_{(1^{m+1})} \,, \quad m = 0, 1, 2, \ldots . 
\label{eq:Weyl-sym-C}
\end{align}
As discussed above, they are equally applicable for surface exponents. We thus obtain, in addition to $x_{(1^q)}^{(s)}$, another sequence of analytically known exponents,
\begin{equation}
   x^{(s)}_{(2,1^m)} = x^{(s)}_{(1^{m+1})} =  
   x_{m+1}^{{\rm h}(s)} = \frac{(2m+1)(m+1)}{3} \,, 
\end{equation}
with $m=0, 1, 2, \ldots$. 

The following comment is in order at this point. The SQH transition as described by the SU(2) network considered here corresponds to a transition between the phases with zero and one edge modes. In principle, one can also consider a situation with a certain integer number of additional ballistic edge modes (in analogy with IQH transitions between higher plateaus). While bulk exponents are insensitive to these additional edge modes, the surface critical behavior will be in general modified \cite{bondesan2012exact}.

\subsubsection{Surface generalized multifractality exponents: Numerical results}
\label{sec:class-C-numerics}

Class C is a class with a (pseudo-)spin degree of freedom, so that pure-scaling eigenfunction observables are given by Eqs.~\eqref{withspin},  \eqref{eq:abelian}.
 As a microscopic model, we use the SU(2) generalization of the Chalker-Coddington random network model \cite{chalker1988percolation,klesse1995universal,kramer2005random,puschmann2021quartic}, which belongs to class C. Wave functions are defined on (directed) links of the network, which form a square lattice. Each random realization of a network is characterized by a unitary evolution matrix, which is built out of matrices on links and at nodes. For each link, we have an SU(2) matrix (chosen randomly, with the Haar measure). The nodes can be subdivided in two sublattices, with scattering matrices having respectively the following form,
\begin{align}
 & S=\mathbb{I}_{2}\otimes\left(\begin{array}{cc}
-\cos\theta & \sin\theta\\
\sin\theta & \cos\theta
\end{array}\right), \nonumber
\\
 & S^{\prime}=\mathbb{I}_{2}\otimes\left(\begin{array}{cc}
-\sin\theta & \cos\theta\\
\cos\theta & \sin\theta
\end{array}\right),
\end{align}
where the factor  $\mathbb{I}_2$ corresponds to the spin SU(2) space. Each link connects two nodes of different sublattices. We are interested here in properties at the critical point, so that we set $\theta=\frac{\pi}{4}$.
We introduce a boundary to this system by setting $\theta=0$ for
all nodes which lie on a straight line going through $S'$ nodes (at $45^\circ$ with respect to the links). 

Numerical results for the polynomial observables 
$\left\langle P_{\lambda}[\psi]\right\rangle$ 
with $|\lambda| \le 4$ are shown in Appendix \ref{sec:scal}, see Fig. \ref{Fig:C_obs}. The analytically predicted scaling reads
\begin{align}
 & L^{(2+\mu)|\lambda|}\left\langle P_{\lambda}[\psi](L,r)\right\rangle 
 r^{\Delta^{(s)}_{(q_1)}+\ldots+\Delta^{(s)}_{(q_n)}} 
 \sim (L/r)^{-\Delta_{\left(q_{1},q_{2},\ldots\right)}^{(s)}},
\end{align}
which differs from  Eq.~\eqref{power_law_sp} by the presence of a non-zero $\mu$ related to the average LDOS scaling.
By plotting the data with $r = 2,3, \ldots, 11$ in this way, we indeed observe a nice collapse. As for class AII above, the straight lines on log-log scale correspond to expected power-law dependence. Further, the fact that we have fans of lines with different slopes confirms that 
$P_{\lambda}[\psi]$ are pure-scaling observables. 
In  Table \ref{table:C_obs} we presented the results for the numerically obtained (by using the data with $r=2$) surface generalized-multifractality exponents. We also included in the table exact analytical values of the exponents $x_{\lambda,{\rm perc}}^{(s)}$  and $\tau_{\lambda,{\rm perc}}^{(s)}$
obtained by mapping to percolation. A very good agreement between the numerical and analytical results is observed.
The Weyl symmetries \eqref{eq:Weyl-sym-C}  are also nicely fulfilled.

In the upper left panel of Fig.~\ref{Fig:C_q_dep} we display the $q$ dependence of critical exponents  $x_{((q/n)^n)}$ with $n=1$, 2, 3, and 4. Dashed lines represent the generalized parabolicity ansatz (\ref{eq:gener-parab}) with $b=1/6$. 
This value of $b$ is chosen in such a way that the analytically known exponents $x_{(1)}^{(s)} = x_{(2)}^{(s)} = 1/3$ are reproduced exactly. It is seen that the generalized parabolicity is strongly violated. In fact, this follows already from the values of analytically known exponents $x_{\lambda,{\rm perc}}^{(s)}$ given in Table~\ref{table:C_obs}.

\begin{table}
 \begin{tabular}{|c||c|c|}
\hline  & $\pi\frac{dx^{(s)}_{(q^{n})}}{dq}\Bigg|_{q=0}$ & $2M \sum_{i=1}^{n}\mathcal{L}_{i}$\\
\hline\hline $n=1$ & $1.805 \pm 0.007$  & $1.821 \pm 0.017$\\
\hline $n=2$ & $8.60 \pm 0.02$ & $8.46 \pm 0.05$\\
\hline $n=3$ & $19.51 \pm 0.08$ & $19.81 \pm 0.19$\\
\hline $n=4$ & $32.84 \pm 0.09$ & $35.54 \pm 0.43$
\\\hline \end{tabular}
\caption{Numerical results for the expressions on both sides  of Eq.~\eqref{1D-2D} (with $n=1$, 2, 3, and 4) at the SQH transition: 
the derivative of the exponent $x^{(s)}_{(q^n)}$ and the sum of $n$ lowest Lyapunov exponents $\mathcal{L}_i$.}
\label{table:C_comparison}
\end{table}

\subsubsection{Lyapunov exponents } 

Implementing a transfer-matrix analysis for the SU(2) network model, we determine Lyapunov exponents. The results for the sums of Lyapunov exponents, up to $n=4$
and  for various values of width $M$, are shown in Fig.~\ref{Fig:C_q_dep}. To check the validity of the exponential-map relation (\ref{1D-2D}), we compare in Table \ref{table:C_comparison} numerical values of the corresponding expressions. It is seen that Eq.~(\ref{1D-2D})  holds with a very good accuracy (corresponding to statistical error bars) for $n=1$, 2, and 3.  Crucially, Eq.~(\ref{1D-2D}) holds with an excellent accuracy ($\sim 1 - 1.5 \%$) not only for $n=1$ but also for $n=2$ and $n=3$. At the same time, the generalized parabolicity is very strongly violated for the corresponding exponents (compare full and dashed lines in the upper left panel of Fig.~\ref{Fig:C_q_dep}). We consider this as a strong evidence in favor of exactness of Eq.~(\ref{1D-2D}), i.e., of invariance with respect to the exponential map.

At the same time, for $n=4$, we observe a somewhat larger deviation. While the difference is still relatively small ($\sim 8\%$), it is a few times larger than expected from statistical errors). While we cannot exclude a possibility that this is a manifestation of weak violation of invariance with respect to the exponential map, we find unlikely that they start to show up only starting from $n=4$. In our view, a much more likely reason is systematic errors due to finite-size effects, which are expected to become stronger for large $n$.

\subsection{Class A: \ IQH transition}
\label{sec:a}

Finally, we turn to the numerical analysis of the IQH transition (class A), which is arguably the most celebrated 2D Anderson-localization critical point. Much work over the past few decades was devoted to attempts to provide an ``educated guess'' for the corresponding field theory at criticality, within the assumption that this is some known CFT, including such closely related theories as free-boson models, Liouville theory, and Wess-Zumino-Novikov-Witten (WZNW) models
\cite{zirnbauer1994towards,zirnbauer1997toward,
janssen1999point-contact,
zirnbauer1999conformal,
kettemann1999information,
tsvelik2001wave,
bhaseen2000towards,
tsvelik2007evidence,
bondesan2017gaussian,zirnbauer2019integer}.
It was found numerically that spectrum of conventional multifractal dimensions $\Delta^{(b)}_{(q)}$ (which is the same as $x^{(b)}_{(q)}$ for class A) is close to parabolic \cite{evers2001multifractality}. Would this parabolicity  be exact, this would provide a support to the above class of theories. However, later works, with improved numerical precision, reported clear (although relatively small) deviations from parabolicity \cite{Evers2008surf,Obuse-Boundary-2008}. Furthermore, these papers also explored the surface multifractality at the IQH transition. It was found that, first, the non-parabolicity is stronger at the surface and, second, the bulk and surface exponents and not related in a simple way. In particular, the ratio $x^{(s)}_{(q)} / x_{(q)}^{(b)}$ is very different from two, which is the value expected for free-boson and related CFTs. 
These results provided a clear numerical evidence against a CFT description of the IQH transition.

We present now  results of our analysis of the surface generalized multifractality at the IQH critical point, which provide additional important insights into the physics of this transition. As for other critical points, our goals include answering the following questions: Does the NL$\sigma$M prediction for pure-scaling observables hold on the surface? Do the Weyl symmetries predicted by the NL$\sigma$M hold? Does the generalized parabolicity hold? Do the relations \eqref{1D-2D} obtained from an assumption of the invariance under the exponential mapping between 2D quasi-1D systems hold?
The class-A Weyl symmetries imply, in particular, the following relations between exponents of polynomial observables:
\begin{align}
 & x_{(1)}=x_{(2,1)}=x_{(2,2)}=0, \nonumber\\
 & x_{(2,1,1)}=x_{(2,2,1)}, \qquad x_{(3,1)}=x_{(3,2)}.
 \label{eq:Weyl-sym-class-A}
\end{align}

\begin{table}
\begin{tabular}{|c||c|c|c|}
\hline $\lambda$ & $\tau_{\lambda}^{(s)}$ & $x_{\lambda}^{(s)}$  & $x_{\lambda}^{(s)}/x_{\lambda}^{(b)}$\\
\hline\hline $\left(1\right)$ & 1.001 & $0.001 \pm 0.004$ & - \\
\hline $\left(2\right)$ & 2.18 & -$0.82 \pm 0.05$ & 1.51 $\pm$ 0.09\\
\hline $\left(1,1\right)$ & 3.94 & $0.94 \pm 0.03$ & 1.66 $\pm$ 0.05\\
\hline $\left(3\right)$ & 2.84 & -$2.16 \pm 0.17$ & 1.30 $\pm$ 0.11\\
\hline $\left(2,1\right)$ & 5.21 & $0.21 \pm 0.08$ & - \\
\hline $\left(1,1,1\right)$ & 7.81 & $2.81 \pm 0.06$ & 1.75 $\pm$ 0.03\\
\hline $\left(4\right)$ & 3.5 & -$3.5 \pm 0.3$ & 1.12 $\pm$ 0.10\\
\hline $\left(3,1\right)$ & 6.15 & -$0.85 \pm 0.17$ & 0.77 $\pm$ 0.16\\
\hline $\left(2,2\right)$ & 7.52 & $0.52 \pm 0.19$ & -\\
\hline $\left(2,1,1\right)$ & 9.27 & $2.27 \pm 0.10$ & 2.06 $\pm$ 0.09\\
\hline $\left(1,1,1,1\right)$ & 12.54 & $5.54 \pm 0.19$ & 1.77 $\pm$ 0.06\\
\hline $\left(3,2\right)$ & 8.68 & -$0.32 \pm 0.24$ & -\\
\hline $\left(2,2,1\right)$ & 11.68 & $2.68 \pm 0.16$ & -
\\\hline \end{tabular}
\caption
{Surface generalized-multifractality critical exponents $\tau_{\lambda}^{(s)}$ and $x_\lambda^{(s)}$ at the IQH transition for all polynomial pure-scaling observables with $\abs{\lambda}\le4$ (and for two observables with $\abs{\lambda}=5$). Statistical error bars (one standard deviation) are shown. The averaging is performed over $2L$ points on the boundary and over $N = 10^4$ realizations
of disorder. The bulk exponents $x^{(b)}_{\lambda}$ are taken from Ref.~\cite{Karcher2021}. The column $x_{\lambda}^{(s)}/x_{\lambda}^{(b)}$ demonstrates independence of surface exponents on bulk ones: $x_{\lambda}^{(s)}/x_{\lambda}^{(b)}\ne \text{const}$.
}
\label{table:A_obs}
\end{table}

\subsubsection{Generalized surface multifractality}

For the numerical analysis, we use the U(1) Chalker-Coddington network model. The difference with respect to the SQH transition study described above 
(Sec.~\ref{sec:class-C-numerics})
is that we do not have a spin degree of freedom now, and the random SU(2) matrices on links are replaced by random U(1) phases.  The boundary is introduced in
the same way as for the SQH transition. For the observables, we use  Eq.~\eqref{eq:abelian} in combination with Eq.~\eqref{withoutspin}, as appropriate for spinless symmetry classes.

Numerical results for the polynomial observables $\left\langle P_{\lambda}[\psi]\right\rangle$ 
are presented in Fig.~\ref{Fig:A_obs}  of Appendix~\ref{sec:scal}. As above, the data is plotted
to verify the scaling prediction \eqref{power_law_sp} and to extract the corresponding exponents $x_{\lambda}^{(s)}$. We observe an already familiar fan of straight lines (on the log-log scale), confirming the predicted form of pure-scaling observables. 
 
Numerically obtained values of surface generalized-multifractality  exponents are presented in  Table \ref{table:A_obs}. It is seen that the Weyl symmetries  \eqref{eq:Weyl-sym-class-A} hold rather well, with deviations of the order of two standard deviations of statistical error bars. It is worth reiterating that higher-order observables are more strongly affected by the limited statistics, which contributes to an apparent slight violation of Weyl relations.

In the bottom left panel of Fig.~\ref{Fig:C_q_dep}, we present results for $q$-dependence of critical exponents $x_{((q/n)^n)}^{(s)}$ for $n\leq 4$.
Dashed lines represent the generalized parabolicity ansatz (\ref{eq:gener-parab}) with $b$ chosen to optimize the fit to $n=1$ data at not too large $q$. 
It is seen that, also for this critical point, the generalized parabolicity is clearly violated. It is worth emphasizing that this violation is evident already for small $q$, where the accuracy of numerical determination of the exponents is very high (within $\sim 1\%$).

\subsubsection{Lyapunov exponents}  

Proceeding in analogy with other critical points, we implement the transfer-matrix analysis for the U(1) Chalker-Coddington network of the IQH transition.
The results for the sums of Lyapunov exponents are shown in Fig.~\ref{Fig:C_q_dep}. In Table
\ref{table:A_comparison} we present a comparison of both sides of the relations \eqref{1D-2D} for $n=1$, 2, 3, and 4. It is seen that the relation holds with an excellent precision (expected on the basis of statistical error bars) for $n=1$, 2, and 3. For $n=4$, we observe a somewhat larger deviation, similarly to the case of SQH transition. However, even though the deviation for $n=4$ substantially exceeds the statistical error bars, it is still quite small numerically ($\sim 3\%$). A plausible reason for an enhanced deviation for $n=4$ was discussed above in the context of SQH critical point.

\begin{table}[h]
 \begin{tabular}{|c||c|c|}
\hline  & $\pi\frac{dx_{(q^{n})}^{(s)}}{dq}\Bigg|_{q=0}$ & $2M\cdot\sum_{i=1}^{n}\mathcal{L}_{i}$\\
\hline\hline $n=1$ & $1.22 \pm 0.01$ & $1.22 \pm 0.01$\\
\hline $n=2$ & $6.18 \pm 0.03$ & $6.20 \pm 0.01$\\
\hline $n=3$ & $14.77 \pm 0.05$ & $14.95 \pm 0.04$\\
\hline $n=4$ & $26.45 \pm 0.08$ & $27.31 \pm 0.09$
\\\hline \end{tabular}
\caption{IQH network: Numerical results for the derivative of subleading exponent $x_{(q^n)}^{(s)}$ and $n$ smallest Lyapunov exponents $\mathcal{L}_i$.}
\label{table:A_comparison}
\end{table}

\subsection{Higher Lyapunov exponents and violation of generalized parabolicity}

As we have seen above, the relation \eqref{1D-2D} between Lyapunov exponents and the $q\to 0$ behavior of generalized-multifractality exponents, which can be reformulated as
\begin{align}
 & 2M \mathcal{L}_n =
 \pi \left (\frac{dx_{(q^{n})}^{(s)}}{dq} - \frac{dx_{(q^{n-1})}^{(s)}}{dq} \right) \! \Bigg|_{q=0}, \ \ n = 1,2, \ldots.
 \label{1D-2D-a}
\end{align}
holds numerically with excellent accuracy and is likely exact. At the same time, the small-$q$ behavior of the exponents demonstrate the strong violation of generalized parabolicity in a very clear form, see Figures \ref{Fig:AII_MIT_q_dep} and \ref{Fig:C_q_dep}. (Importantly, the $q\to 0$ limit corresponds to averaging of logarithm (or, equivalently, to typical values), implying a particularly high numerical accuracy.) Thus, the sequence of Lyapunov exponents by itself can serve as a smoking gun for violation of generalized parabolicity (and thus of conformal invariance). In Table \ref{table:Lyapunov_exp} (Appendix \ref{sec:Lyapunov}) we summarize numerical results  for the Lyapunov exponents $\mathcal{L}_n$ with $n=1$, 2, 3, and 4 for the critical points studied in this paper. 

 If the multifractality spectrum obeyed generalized parabolicity, the sequence of Lyapunov exponents would be proportional to $-c_n$ [for the corresponding symmetry class, see Eq.~\eqref{eq:cj-A-AII-C}]. This indeed holds very well for the class-AII metal. At the same time, this property is strongly violated for all three Anderson-transition critical points (class-AII MIT as well as SQH and IQH transitions), demonstrating strong violation of generalized parabolicity. Interestingly, the Lyapunov exponents are nevertheless equidistant with a good accuracy. For example, for the AII MIT we find numerically ${\cal L}_1 : {\cal L}_2 : {\cal L}_3 : {\cal L}_4 \approx 1 : 3 : 5 :7$, to be compared with the sequence $1:5:9:13$ that would correspond to generaliized parabolicity.

 For comlpeteness, we also included Table \ref{table:Lyapunov_exp_bulk} that contains results for Lyapunov exponents ${\cal L}^{(p)}_n$ evaluated at AII MIT, SQH, and IQH transitions for strips with periodic boundary conditions, which correspond to bulk generalized multifractality, see Eq.~\eqref{1D-2D-bulk}. Violation of generalized parabolicity is evident also in this case, as well as the fact that there is no simple relation between the exponents ${\cal L}_n$ and ${\cal L}^{(p)}_n$ (or, equivalently, between surface and bulk generalized multifractality) at strong-coupling criticality. (Note that for a metal, the Lyapunov exponents are, to the leading order, independent on boundary conditions, in correspondence with Eq.~\eqref{eq:xs-xb}.)

\section{Summary and outlook}
\label{sec:summary}

In this paper, we have extended the concept of generalized multifractality to boundary of a system at Anderson-transition criticality. We have further numerically explored generalized surface multiftactality in 2D systems of classes AII (at the MIT and in the``weakly critical'' metallic phase), C (at the SQH transition), and A (at the IQH transition). We have verified that Eqs.~(\ref{withoutspin}), (\ref{withspin}), and \eqref{eq:abelian} correctly give pure-scaling observables also on the system surface and extracted numerical values of surface generalized-multifractality exponents. The exponents obey well the Weil symmetry relations. These results confirm the validity of analytical predictions based on NL$\sigma$M. 

Deeply in the metallic phase of class AII, the exponents satisfy with a good acurracy the generalized parabolicity \eqref{eq:gener-parab}, as expected in the one-loop approximation that is parametrically justified in this regime. Further, the surface exponents are approximately equal to twice the bulk exponents in this regime, $x_{\lambda}^{(s)}=2 x_{\lambda}^{(b)}$, again in agreement with analytical predictions. 

On the other hand, the generalized parabolicity \eqref{eq:gener-parab} is strongly violated at the MIT of class AII as well as at SQH and IQH transitions. Further, there is no simple relation between surface and bulk exponents at these critical points. This corroborates earlier conclusions that these critical points are not described by a theory of WZNW or free-boson type and, more generally, not described by a CFT. 

We have further derived relations \eqref{1D-2D} that hold under an assumption that the theory is invariant under the exponential map \eqref{eq:expm} between a quasi-1D and a 2D system. This formula connects {\it typical} values of the generalized-multifractality observables $P_{(1^n)}[\psi]$ in a 2D system with Lyapunov exponents of a quasi-1D system. We put the relations \eqref{1D-2D} under scrutiny in different symmetry classes A, AII and C. In class AII, we analyze both the thermal metal phase and the metal-insulator transition. Our results show that the relations \eqref{1D-2D} hold with an excellent accuracy and are most likely exact. (Deviations that substantially exceed statistical error bars are observed only in the case of $n=4$ for the SQH and IQH transition, and even they are of the order of a few percent. We attribute these deviations to an increase of finite-size corrections at large $n$.) Thus, we interpret the obtained results as a strong evidence of invariance of the theory with respect to the exponential map, at least for this class of observables. 

We close the paper by briefly discussing open questions and associated prospects for future research. 

\begin{itemize}

\item A natural perspective is extending of our numerical analysis of surface generalized-multifractality to 2D Anderson-transition critical points of other symmetry classes. Also, we foresee a numerical study of generalized multifractality in system dimensionality $d>2$ (in particular, at the conventional Anderson transition in class AI in three dimensions).

\item Our results, in combination with earlier findings, indicate that 2D Anderson-transition critical points have very peculiar properties. They do not possess full conformal invariance but at the same exhibit invariance with respect to some conformal transformations (in addition to scale invariance)---at least with respect to the exponential map \eqref{eq:expm}. It would be interesting to understand better under what classes of conformal transformation such systems are invariant (respectively, not invariant). A further challenging task is to extend this analysis to systems above two dimensions and to explore whether there are some analogies in a broader context of critical phenomena, see in this relation Ref.~\cite{padayasi2023conformal}.

\item Much previous work was devoted to investigation of the interplay of multifractality (conventional and generalized)  in the bulk and inter-particle (electron-electron) interaction. This research area has many facets. In particular, it includes effects of multifractality on dynamical scaling at Anderson transitions 
\cite{lee1996effects,wang2000short-range,burmistrov2011wave}, influence of Coulomb interaction on spectra of multifractal exponents \cite{burmistrov2013multifractality,Amini_2014,Burmistrov2015multi,burmistrov2016mesoscopic,Lee2018,Babkin2022}, and instabilities induced \cite{foster2012interaction,foster2014topological} or enhanced by multifractality. A remarkable manifestation of the last point is a parametric enhancement of superconducting transition temperature by multifractality in 2D systems \cite{burmistrov2012enhancement,DellAnna2013,burmistrov2015superconductor,Gastiasoro2018,fan2020enhanced,Stosiek2020,burmistrov2021multifractallyenhanced,Andriyakhina2022} (and also near Anderson transition in 3D systems \cite{feigel2007eigenfunction,feigel2010fractal,mayoh2015global,garcia-garcia2020superconductivity}); recent experimental works \cite{zhao2019disorder-induced,rubio-verdu2020visualization} reported observation of this effect in monolayer niobium dichalcogenides. Extending the research in these directions to the interplay of (generalized) multifractality and interaction at a surface would be of much interest.

\end{itemize}


\section*{Acknowledgement}
We thank Ilya Gruzberg for many illuminating discussions. S.S.B., J.F.K., and A.D.M. acknowledge support by the Deutsche Forschungsgemeinschaft (DFG) via the grant MI 658/14-1. I.S.B. acknowledges support from Russian Science Foundation (grant No. 22-42-04416). 

\onecolumngrid

\appendix 
\section{Numerical verification of pure scaling observables and determination of scaling exponents}
\label{sec:scal}  

In this appendix, we show numerical data for the scaling of observables $P_\lambda[\psi]$, as given by Eq.~\eqref{eq:abelian} in combination with either Eq.~\eqref{withspin} (classes AII and C) or Eq.~\eqref{withoutspin} (class A) for the metallic systems of class AII and for critical systems at SQH and IQH transitions. The data support the claim that  $P_\lambda[\psi]$ exhibit pure-scaling behavior and are used to extract the values of surface generalized-multifractality exponents presented in the main text. All the data in this appendix is presented in a form analogous to Fig.~\ref{Fig:AII_MIT_obs} of the main text (where results for the class-AII MIT are shown).

\begin{figure*}[h]
\center
\includegraphics[width=0.32\textwidth]{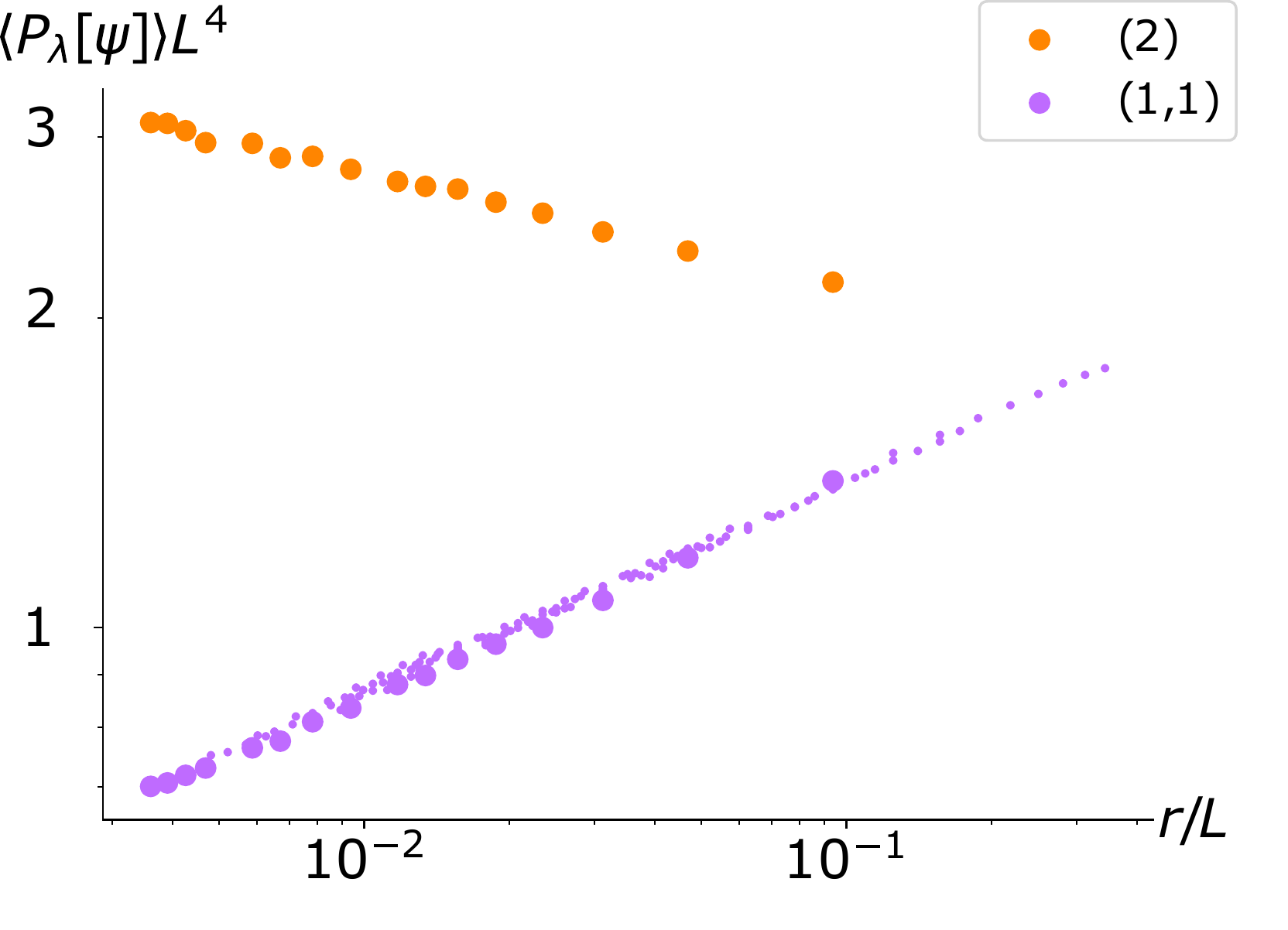}
\includegraphics[width=0.32\textwidth]{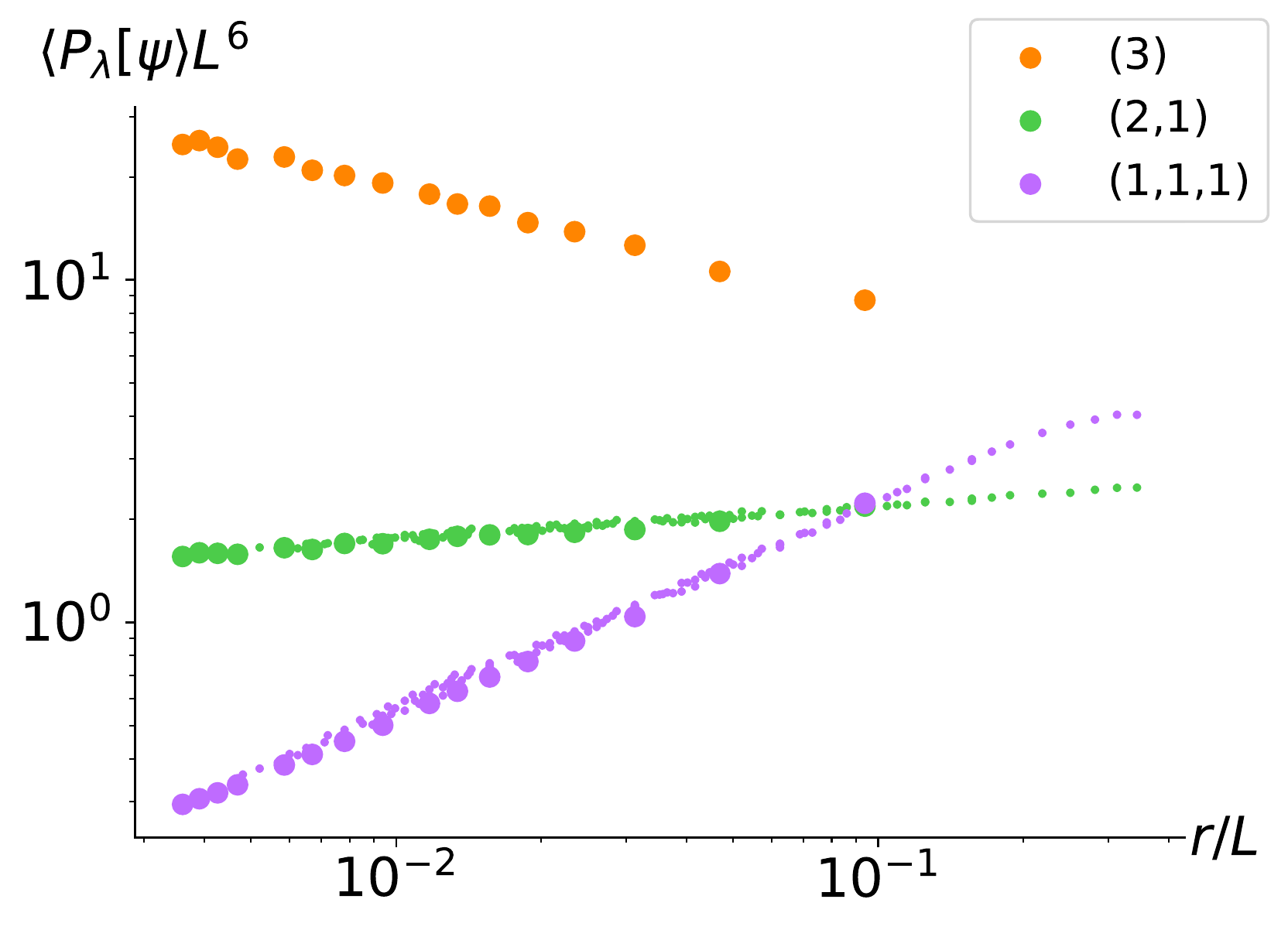} \includegraphics[width=0.32\textwidth]{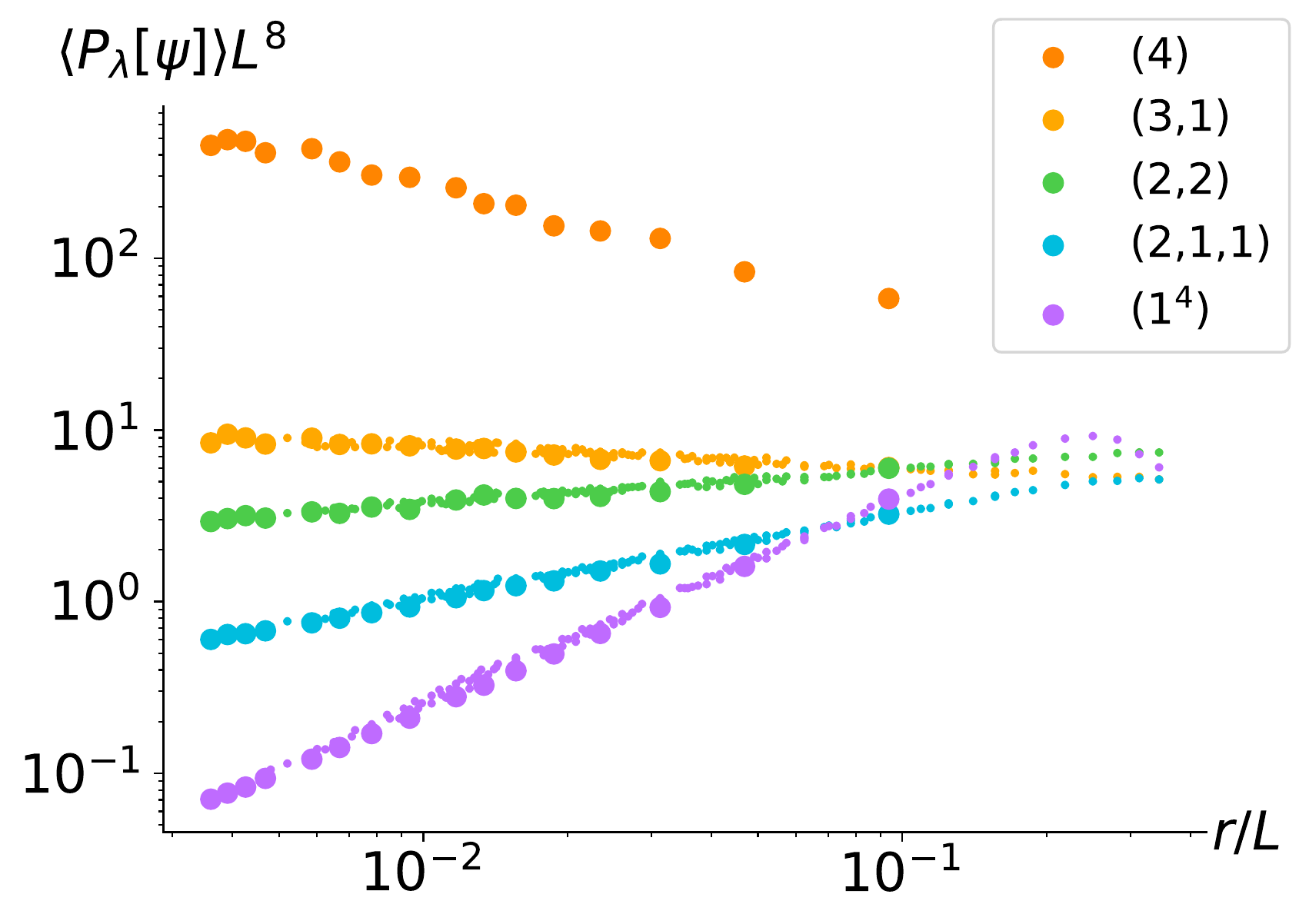}
\caption
{Generalized multifractality in the metallic phase of  class AII (Ando model with $W=3$) for polynomial observables with $|\lambda|=2$ (left), $3$ (middle), and $4$ (right). The pure-scaling observables $ L^{2\left|\lambda\right|} \langle P_{\lambda}[\psi](L,r) \rangle$
are averaged over $N=10^4$ realizations of disorder and over points on the boundary. The data is scaled with  $r^{\Delta_{(q_1)}+...+\Delta_{(q_n)}}$, yielding a collapse as a function of $r/L$ as predicted. Data corresponding to the smallest $r=3$ is highlighted as large dots.
}  
\label{Fig:AII_metal_obs}
\end{figure*}

\begin{figure*}[h]
\centering
\includegraphics[width=0.32\textwidth]{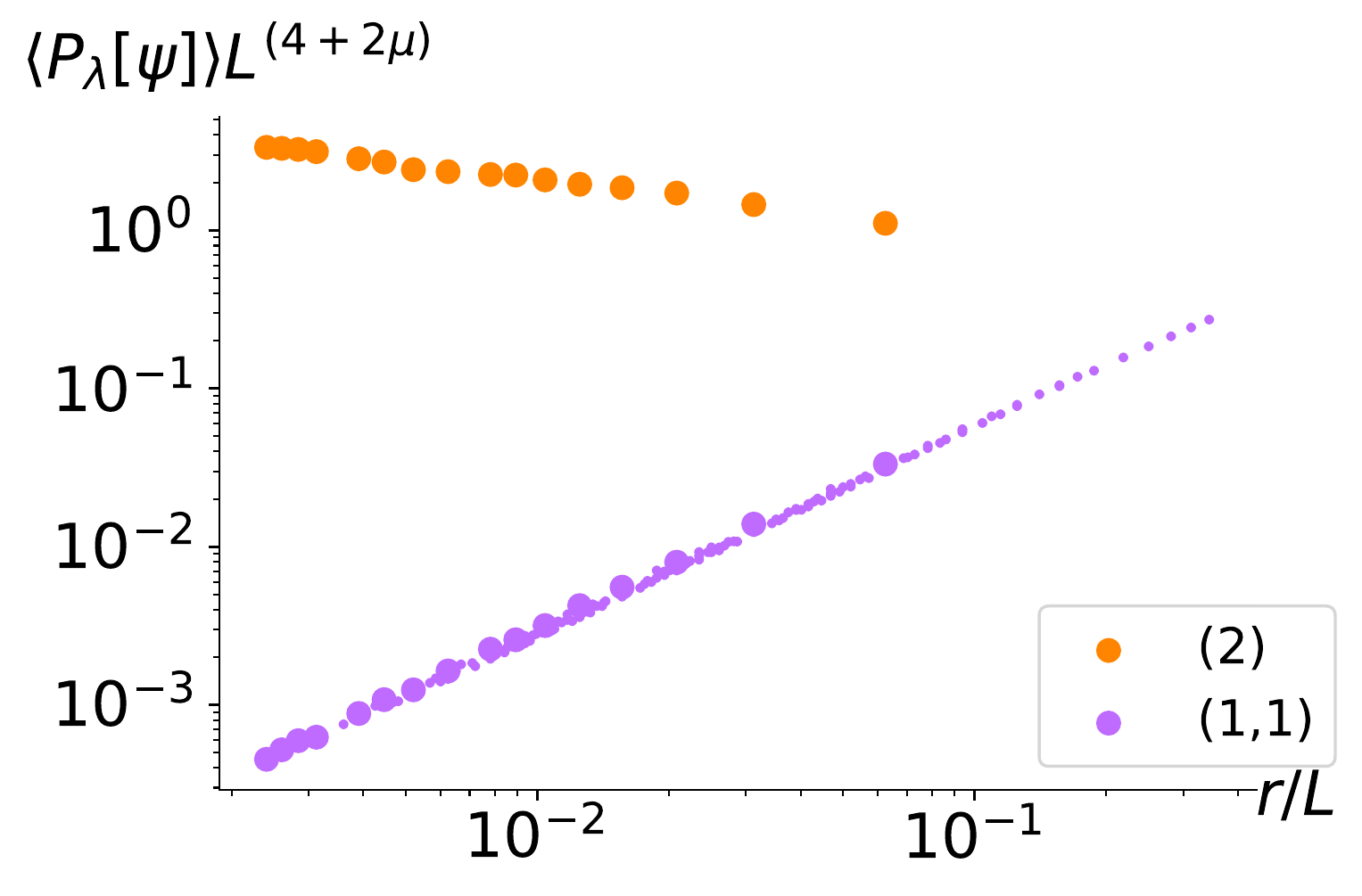}
\includegraphics[width=0.32\textwidth]{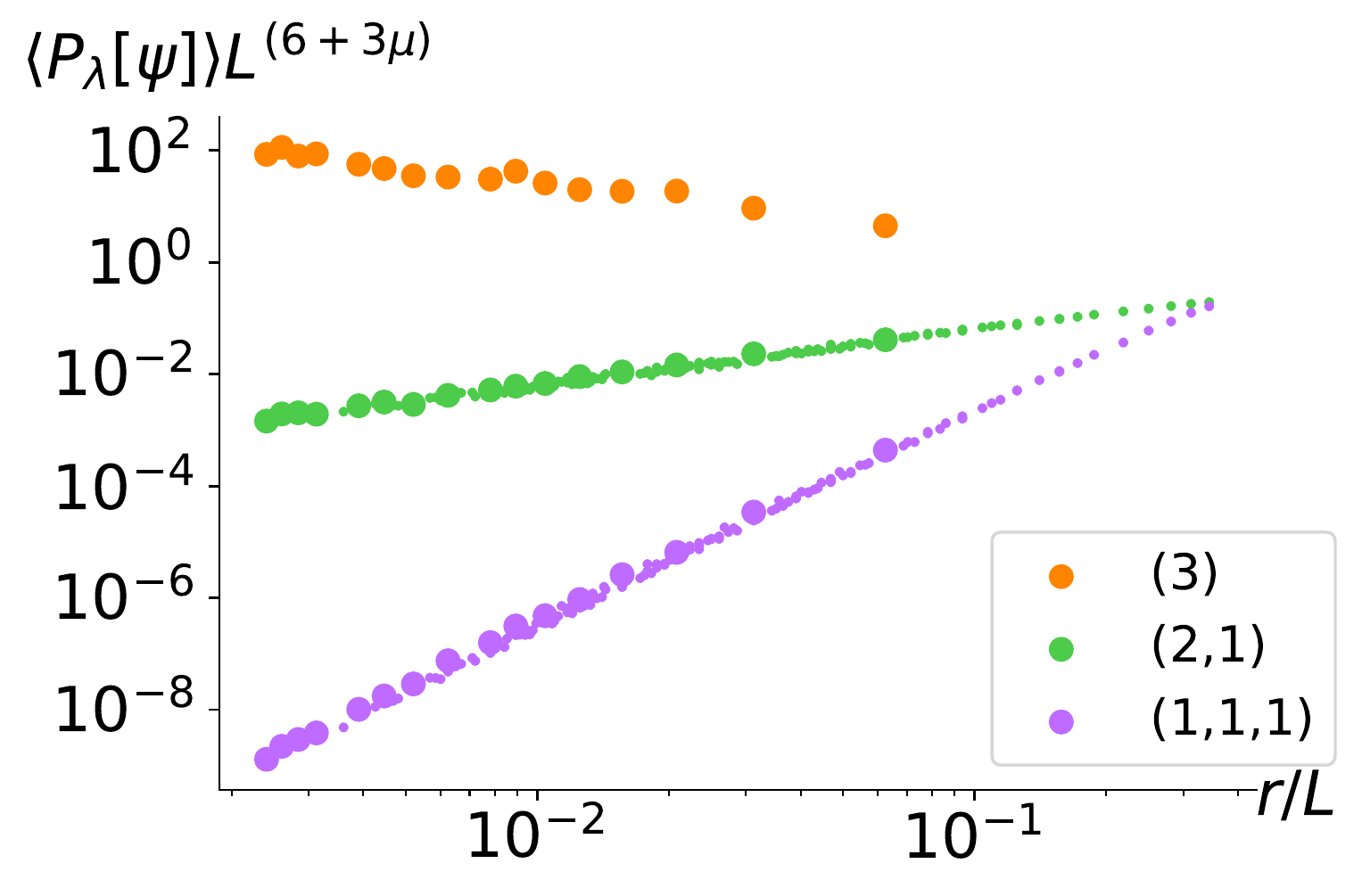}
\includegraphics[width=0.32\textwidth]{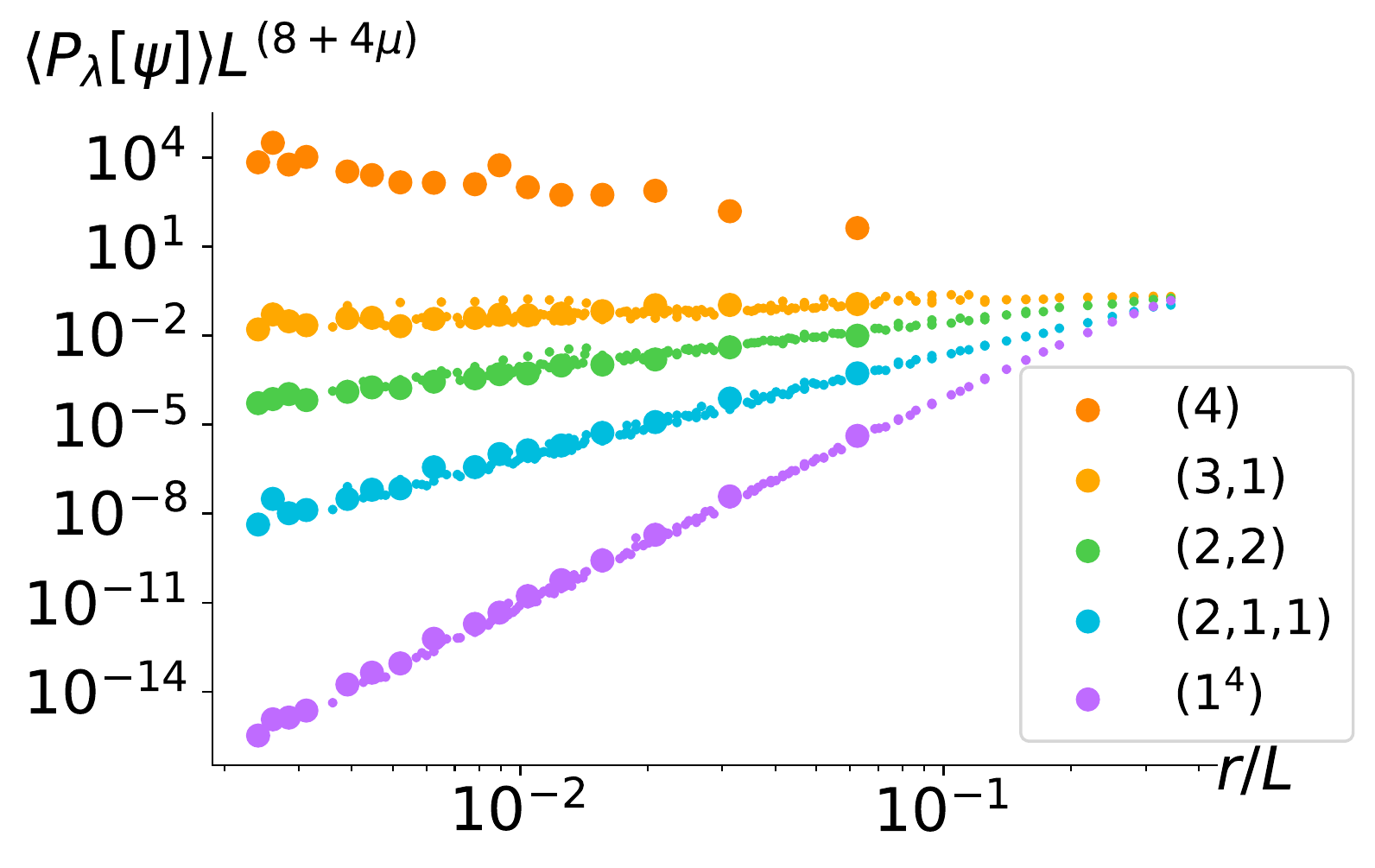}
\caption
{Generalized multifractality at SQH transiton (class C) for polynomial observables with $|\lambda|=2$ (left), $3$ (middle), and $4$ (right). The pure-scaling observables $L^{(2+\mu) \left|\lambda\right|} \langle P_{\lambda}[\psi](L,r) \rangle$
are averaged over $N=10^4$ realizations of disorder and over points on the boundary. The data is scaled with  $r^{\Delta_{(q_1)}+...+\Delta_{(q_n)}}$, yielding a collapse as a function of $r/L$ as predicted. Data corresponding to the smallest $r=2$ is highlighted as large dots.
} 
\label{Fig:C_obs}
\end{figure*}

\begin{figure*}[h]
\centering
\includegraphics[width=0.32\textwidth]{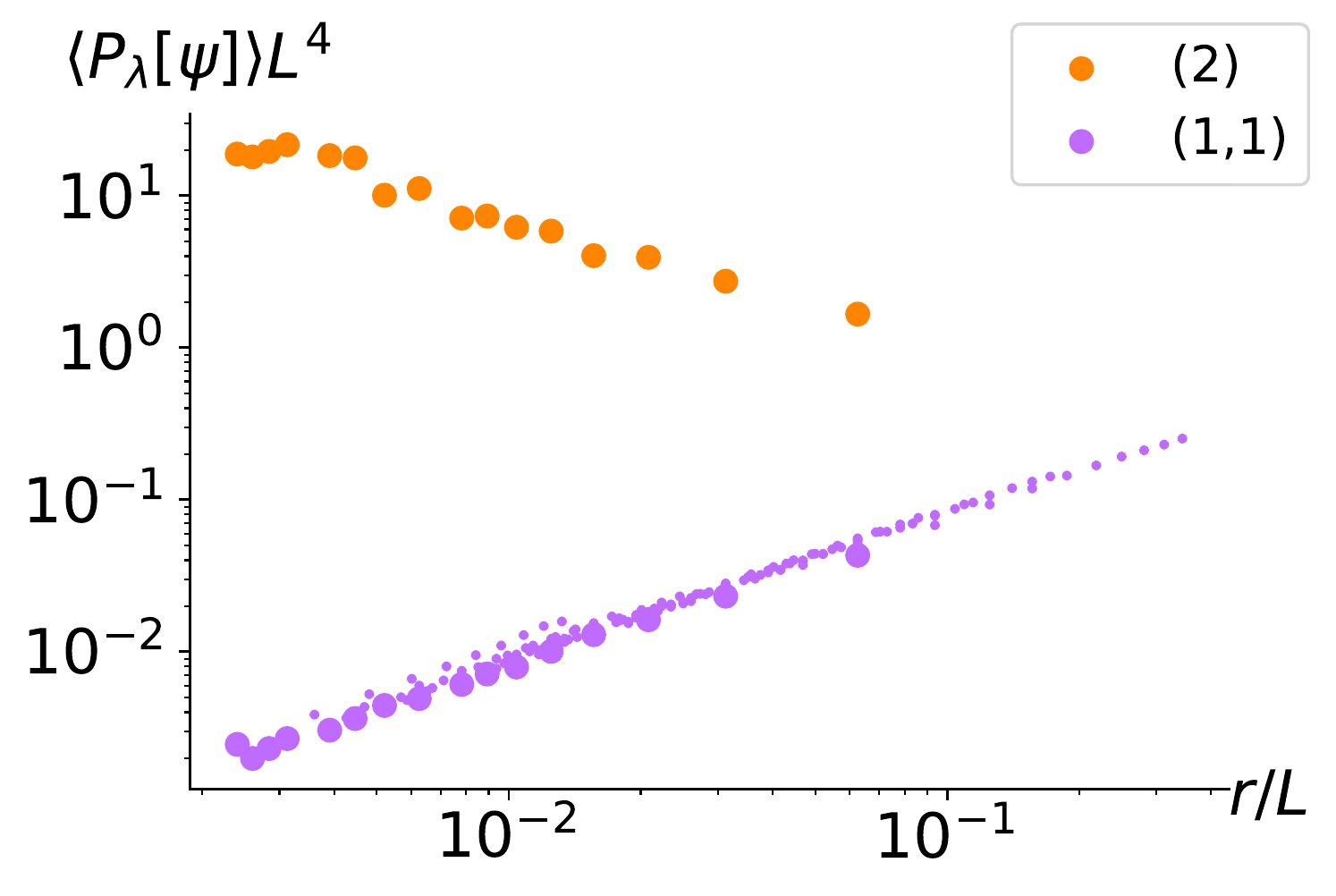}
\includegraphics[width=0.32\textwidth]{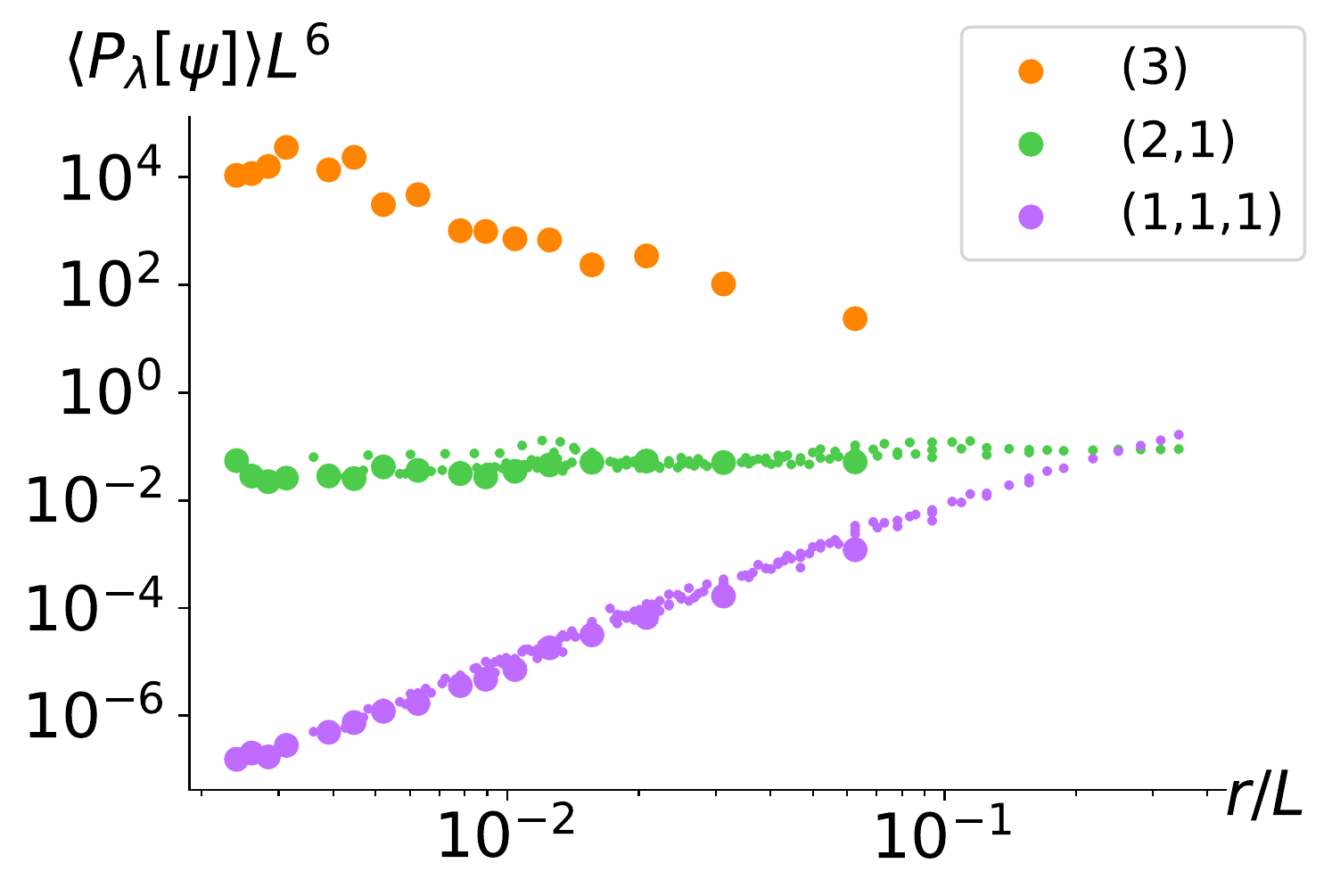}
\includegraphics[width=0.32\textwidth]{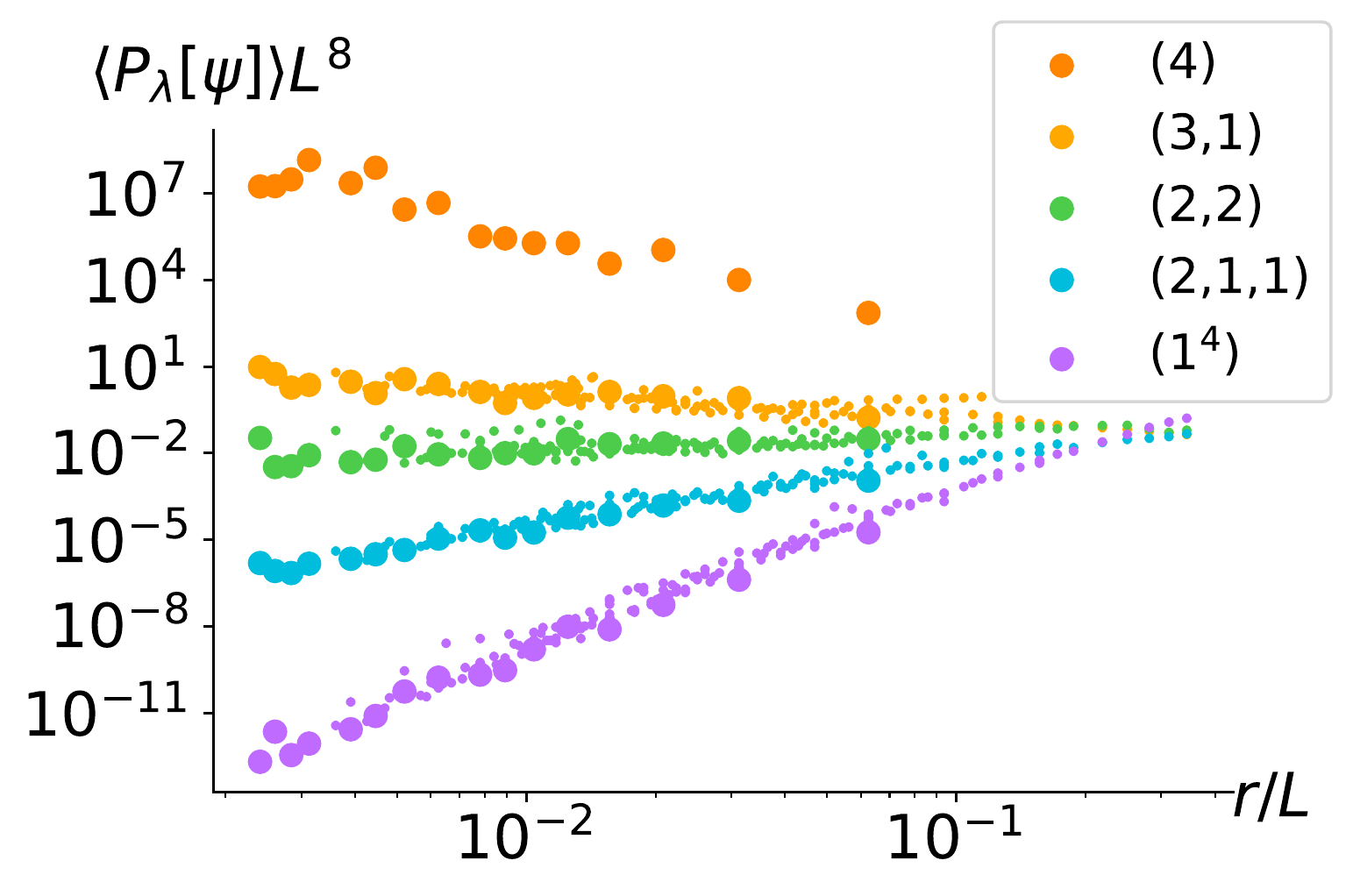}
\caption
{Generalized multifractality at IQH transition (class A) for polynomial observables with $|\lambda|=2$ (left), $3$ (middle), and $4$ (right). The pure-scaling observables $L^{2\left|\lambda\right|} \langle P_{\lambda}[\psi](L,r) \rangle$
are averaged over $N=10^4$ realizations of disorder and over points on the boundary. The data is scaled with  $r^{\Delta_{(q_1)}+...+\Delta_{(q_n)}}$, yielding a collapse as a function of $r/L$ as predicted. Data corresponding to the smallest $r=2$ is highlighted as large dots.
 }  
\label{Fig:A_obs}  
\end{figure*}

In Fig. \ref{Fig:AII_metal_obs}, the numerical results for the metallic phase of the Ando model in class AII are shown. In Figs. \ref{Fig:C_obs} and \ref{Fig:A_obs}, we present the data for the network models in classes C (SQH transition) and A (IQH transition), respectively.

\section{Lyapunov exponents}
\label{sec:Lyapunov}

In this Appendix, we summarize results for the first  four Lyapunov exponents  for all the critical points studied numerically in this work. The Lyapunov exponents ${\cal L}_n$ (multiplied by $2M$) are presented in Table \ref{table:Lyapunov_exp}. All of them are calculated for strips with open boundary conditions (that correspond to surface generalized multifractality in the framework of exponential map). For completeness, we also present (Table \ref{table:Lyapunov_exp_bulk}) the results for Lyapunov exponents ${\cal L}^{(p)}_n$ evaluated with periodic boundary conditions, which correspond to bulk generalized multifractality. 

\begin{table}[h]
 \begin{tabular}{|c||ccc|ccc|ccc|ccc|}
\hline  
& & AII  MIT & & & AII  metal &  & & SQH &  & & IQH & \\
& $2M\mathcal{L}_{n}$ & $\mathcal{L}_{n}/\mathcal{L}_{1}$ & $- c_n$ & $2M \mathcal{L}_{n}$ & $\mathcal{L}_{n}/\mathcal{L}_{1}$ & $- c_n$ & $2M \mathcal{L}_{n}$ & $3\mathcal{L}_{n}/\mathcal{L}_{1}$ & $- c_n$ & $2M\mathcal{L}_{n}$ & $\mathcal{L}_{n}/\mathcal{L}_{1}$ & $-c_n$\\
\hline\hline $n=1$ & $1.331 \pm 0.005$ & $1$ & 1 & $0.174 \pm 0.005$ &  $ 1$ & 1 & $1.821 \pm 0.017$ & $3$ & 3 & $1.22 \pm 0.01$ &  $ 1$ & 1\\
\hline $n=2$ & $4.062 \pm 0.016$ & $3.05\pm0.02$  &  5 & $0.859\pm 0.007$ & $4.94\pm0.18$  &  5  & $6.64\pm0.04$ & $10.94\pm0.17$  &  7 & $4.95\pm0.01$ & $4.06\pm0.04$  &  3\\
\hline $n=3$ & $6.66\pm0.04$ & $5.00\pm0.05$  & 9 & $1.531\pm0.005$ & $8.8\pm0.3$  &  9  & $11.35\pm0.14$ & $18.7\pm0.4$  &  11 & $8.74\pm0.03$ & $7.16\pm0.08$  &  5\\
\hline $n=4$ & $9.20\pm0.08$ & $6.91\pm0.09$  & 13 & $2.239\pm0.009$ &  $12.9\pm0.4$ &  13  & $15.7\pm0.2$ & $25.9\pm0.6$  &  15 & $12.36\pm0.05$ & $10.13\pm0.12$  &  7
\\\hline \end{tabular}
\caption{Lyapunov exponents $\mathcal{L}_n$ with $n=1$, 2, 3, and 4  (for strips with open boundary conditions)  at critical points studied numerically in this paper. If generalized parabolicity held, the sequence ${\cal L}_n$ would be proportional to $-c_n$. The second and first column for each critical point nicely illustrate that the surface generalized parabolicity holds for the AII metal but is strongly violated at the AII MIT, SQH, and IQH critical points. }
\label{table:Lyapunov_exp}
\end{table}

\begin{table}[h!]
 \begin{tabular}{|c||ccc|ccc|ccc|}
\hline  
& & AII  MIT  &  & & SQH &  & & IQH & \\
& $2M\mathcal{L}_{n}^{(p)}$ & $\mathcal{L}_{n}^{(p)}/\mathcal{L}^{(p)}_{1}$ & $- c_n$ &  $2M \mathcal{L}_n^{(p)}$ & $3\mathcal{L}_{n}^{(p)}/\mathcal{L}_{1}^{(p)}$ & $- c_n$ & $2M\mathcal{L}_{n}^{(p)}$ & $\mathcal{L}_{n}^{(p)}/\mathcal{L}_{1}^{(p)}$ & $-c_n$\\
\hline\hline $n=1$ & $0.533 \pm 0.028$ & $1$ & 1 &  $1.205 \pm 0.032$ & $3$ & 3 & $0.830 \pm 0.012$ &  $ 1$ & 1\\
\hline $n=2$ & $2.00 \pm 0.03$ & $3.74\pm0.20$  &  5   & $3.654\pm0.034$ & $9.10\pm0.25$  &  7 & $2.76\pm0.01$ & $3.32\pm0.05$  &  3\\
\hline $n=3$ & $3.33\pm0.02$ & $6.24\pm0.33$  &  9  & $6.02\pm0.04$ & $15.0\pm0.4$  &  11 & $4.64\pm0.01$ & $5.59 \pm0.09$  &  5\\
\hline $n=4$ & $4.623 \pm0.03$ & $8.7\pm0.5$  &  13  & $8.30\pm0.04$ & $20.6\pm0.5$  &  15 & $6.53\pm0.02$ & $7.86\pm0.12$  &  7
\\\hline \end{tabular}
\caption{Lyapunov exponents $\mathcal{L}_n^{(p)}$ with $n=1$, 2, 3, and 4  (for strips with periodic boundary conditions corresponding to bulk generalized multifactality). If generalized parabolicity held, the sequence ${\cal L}^{(p)}_n$ would be proportional to $-c_n$. The second and first column for each critical point nicely illustrate that the bulk generalized parabolicity is strongly violated at the AII MIT, SQH, and IQH critical points.}
\label{table:Lyapunov_exp_bulk}
\end{table}

\twocolumngrid

\bibliography{gener-MF}
\addcontentsline{toc}{section}{References}

\end{document}